\documentclass[11pt,a4paper]{article}
\usepackage[onehalfspacing]{setspace}
\usepackage[utf8]{inputenc}
\usepackage{amsmath}
\usepackage{amsfonts}
\usepackage{amssymb}
\usepackage{graphicx}
\usepackage{lmodern}
\usepackage[left=2.5cm,right=2.5cm,top=2.5cm,bottom=2.5cm]{geometry}
\usepackage[authoryear]{natbib}
\usepackage{authblk}
\usepackage[colorlinks=true, urlcolor=citecolor, linkcolor=citecolor, citecolor=citecolor]{hyperref}
\usepackage{amsmath}
\usepackage{bm}
\usepackage[caption=false]{subfig}
\usepackage[flushleft]{threeparttable}
\usepackage{float}
\usepackage{autonum} 
\usepackage{helvet}
\usepackage[T1]{fontenc}
\usepackage{fonttable}
\newcommand{\bcx}{{\bf X}}

\newcommand{\bcb}{{\bf B}}

\newcommand{\bcd}{{\bf D}}

\newcommand{\bck}{{\bf K}}
\newcommand{\bcp}{{\bf P}}
\newcommand{\bct}{{\bf T}}
\newcommand{\ba}{{\bf a}}

\newcommand{\bx}{{\bf x}}

\newcommand{\bt}{{\bf t}}

\DeclareMathOperator{\diag}{diag}

\let\proglang=\textsf
\newcommand{\pkg}[1]{{\fontseries{b}\selectfont #1}}
\makeatletter
\newcommand\code{\bgroup\@makeother\_\@makeother\~\@makeother\$\@codex}
\def\@codex#1{{\normalfont\ttfamily\hyphenchar\font=-1 #1}\egroup}
\makeatother
\usepackage{authblk}
\title{Nonlinear association structures in flexible Bayesian additive joint models}
\author[1]{Meike K\"ohler}
\author[2]{Nikolaus Umlauf}
\author[3]{Sonja Greven\thanks{Corresponding author: sonja.greven@stat.uni-muenchen.de}}
\affil[1]{Institute of Diabetes Research, Helmholtz Zentrum M\"unchen, and Forschergruppe Diabetes, Klinikum rechts der Isar, Technische Universit\"at M\"unchen, Neuherberg, Germany}
\affil[2]{Department of Statistics, Faculty of Economics and Statistics, 
Universit\"at Innsbruck, Innsbruck, Austria}
\affil[3]{Department of Statistics, Ludwig-Maximilians-Universit\"at M\"unchen, M\"unchen, Germany}
\date{}                     
\begin{document}

\maketitle

\begin{abstract}
Joint models of longitudinal and survival data have become an important tool for modeling associations between longitudinal biomarkers and event processes. The association between marker and log-hazard is assumed to be linear in existing shared random effects  models, with this assumption usually remaining unchecked. We present an extended framework of flexible additive joint models that allows the estimation of nonlinear, covariate specific associations by making use of Bayesian P-splines. 
Our joint models are estimated in a Bayesian framework using structured additive predictors for all model components, allowing for great flexibility in the specification of smooth nonlinear, time-varying and random effects terms for longitudinal submodel, survival submodel and their association. 
The ability to capture truly linear and nonlinear associations is assessed in simulations and illustrated on the widely studied biomedical data on the rare fatal liver disease primary biliary cirrhosis. All methods are implemented in the \proglang{R} package \pkg{bamlss} to facilitate the application of this flexible joint model in practice.
\end{abstract}

\section{Introduction}
\label{sec1}

The joint modeling of longitudinal and survival processes has gained large attention in the last decade and has seen a broad range of developments. In this work we present a flexible framework for Bayesian additive joint models that allows for a highly flexible specification of the association between a longitudinal biomarker and a survival process to gain further insights into complex diseases. A special focus is placed on potentially nonlinear associations between a longitudinal biomarker and the log-hazard of an event. 

The research into joint models has largely been motivated by biomedical applications such as modeling of CD4 counts and HIV progression \citep{wulfsohn_joint_1997, tsiatis_semiparametric_2001}, PSA values and prostate cancer \citep{taylor_real-time_2013} or breast cancer \citep{chi_joint_2006} and receives growing attention in applied research \citep{sudell_joint_2016}.  In all these applications there is a need for unbiased modeling of a longitudinal covariate, often a biomarker, and its association to the hazard of an event. This situation demands a special treatment as the longitudinal covariate is potentially subject to measurement error, measured at individual-specific timepoints and as an internal time-varying covariate only observed until the occurrence of the event. Joint models take all these complications into account  by formulating a joint likelihood for the longitudinal and the survival submodel and thereby achieve an unbiased modeling of both. As a detailed overview of the field of joint models for longitudinal and time-to-event data is beyond the scope of this work, we refer to the excellent reviews on the topic from \cite{tsiatis_joint_2004}, \cite{rizopoulos_joint_2012} and \cite{gould_joint_2015}. The main idea of this modeling framework is that a set of parameters is assumed to influence both the longitudinal and the survival submodel with conditional independence between the two models, given those parameters. This shared parameter linking the two submodels can be a latent class structure, as in joint latent class models \citep{proust-lima_joint_2014}, or random effects, as is the case in most developments in joint modeling. The associations between longitudinal marker and log-hazard in this class of shared random effects models can be parameterized differently such that only the random effects, the current value of the marker or further transformations of this current value are associated (see \citet{hickey_joint_2016} for an overview of associations structures in multivariate joint modeling). Focus in this work is placed on the current value association. 

Existing shared random effects models include the linearity assumption that the effect of the modeled marker trajectories on the logarithm of the hazard is linear. In the context of survival analysis checking the linearity assumption as well as the modeling of an appropriate functional form has been under study \citep{buchholz_comparison_2011, hollander_estimating_2006}. In different biomedical applications it was shown that appropriate modeling of the functional form of continuous covariate effects reduces bias and allows for additional insights into prognostic factors, for example in the study of breast cancer \citep{gray_flexible_1992, sauerbrei_modelling_1999}, lung cancer \citep{gagnon_flexible_2010} and leukemia \citep{inaba_effect_2012}. For the accurate specification of nonlinear effects of continuous covariates in the time-to-event model different strategies have been applied, such as fractional polynomials \citep{royston_regression_1994, sauerbrei_new_2007} as well as unpenalized \citep{sleeper_regression_1990, wynant_flexible_2016} and penalized spline approaches \citep{hastie_generalized_1995, hofner_building_2011}. 

The results from survival modeling suggest that the linearity assumption may also not always be met when modeling the effects of a longitudinal marker in a joint model. To our knowledge, to date no shared random effects joint model approach extends or even tests this assumption. The user of a joint model can only assume that, given an appropriate transformation of the raw marker values such as a log-transformation, the association is indeed linear. The present work aims to fill this gap by allowing greater flexibility in the specification of the association between marker and event. Note that joint latent class models \citep{proust-lima_joint_2014}, where the latent class is associated with the log-hazard and the association between marker and event is only implicit, also allow for a particular kind of nonlinear relationship between marker and hazard. However our interest lies in gaining insights into the detailed nature of this association, and therefore an explicit modeling of this association is necessary.

We have previously presented a general framework for flexible structured additive joint models \citep{kohler_flexible_2017} with the focus on modeling highly subject-specific nonlinear individual longitudinal trajectories as well as a time-varying association. 
This flexibility is achieved by formulating the joint model as a structured additive regression \citep{fahrmeir_penalized_2004} in which all model parts, which are the baseline hazard, baseline and time-varying covariate effects, mean and variance of the modeled longitudinal marker as well as the association are structured additive predictors. These predictors can encompass nonlinear, smooth and time-varying effects by making use of P-splines \citep{EilersMarx1996} and capture highly flexible nonlinear individual trajectories by modeling them as functional random intercepts \citep{scheipl_functional_2015}. The model is estimated in a Bayesian framework with smoothness and random effects structures induced by appropriate prior specifications. In the present work this framework is generalized further to allow for nonlinear associations between a marker and the event process as well as to allow this nonlinear association to vary with covariates.

In order to facilitate the application of this flexible joint model it is fully implemented in the \proglang{R} package \pkg{bamlss} thereby adding to the available range of joint model packages. Software packages in the shared random effects approach are \pkg{JM} \citep{JM} and its Bayesian counterpart \pkg{JMbayes} \citep{R_JMbayes}, \pkg{joineR} \citep{joineR}, \pkg{frailtypack} \citep{frailtypack} as well as the stata package \pkg{stjm} \citep{crowther_stjm} and the SAS macro \pkg{JMFit} \citep{JMFit} of which many are rather restricted in the amount of flexibility they allow in modeling nonlinear individual trajectories and the association itself. Out of these packages up to date the \proglang{R} package \pkg{JMbayes} offers the most flexibility in modeling individual trajectories and different association structures while, however, also assuming linearity in the association between the marker and the log-hazard. We therefore compare our implementation with this established package in our simulation study. 
 
The paper is structured as follows: Section \ref{sec2} presents the general framework  with details on the Bayesian estimation in Section \ref{sec3}. An extensive simulation study in Section \ref{sec4} and a practical application of the model on the well-known data on primary biliary cirrhosis (PBC) \citep{murtaugh_primary_1994}, which is included in the \proglang{R} package \pkg{JMbayes}, in Section \ref{sec5}  give further insights into the performance of this flexible model. Section \ref{sec6} presents concluding remarks and further technical details can be found in the Appendix. The presented methods are implemented in the \proglang{R} package \pkg{bamlss}. A current version of the package can be found on (\url{https://github.com/Meike-K/bamlss-dev}) and source code to fully reproduce the results of the simulations and the application is given in the Supplementary Information. The developments of this work will be included in the next CRAN update of the \proglang{R} package \pkg{bamlss}.

\section{A flexible additive joint model}
\label{sec2}

In the following we further generalize the previously formulated flexible additive joint model \citep{kohler_flexible_2017} to allow for complex nonlinear association structures between a longitudinal marker and the time-to-event process. 

\subsection{General model}

For each subject $i = 1, \dots, n$ we observe the longitudinal response $\bm{y}_{i}=[y_{i1}, \cdots, y_{i n_i}]^{\top}$ at the potentially subject-specific time points $\bt_i=[t_{i1}, \cdots, t_{i n_i}]^{\top}$  with $t_{i1}\leq \cdots \leq t_{in_{i}}\leq T_{i}$, modeled by 

\begin{equation}
y_{ij}=\eta_{\mu i}(t_{ij})+\varepsilon_{ij} \text{ with } \varepsilon_{ij} \sim N(0, \exp[\eta_{\sigma i}(t_{ij})]^2).
\label{eq:long} 
\end{equation}
The predictor $\eta_{\mu}$ denotes the "true" longitudinal marker that serves as a time-varying covariate in the time-to-event model. Additionally we observe for every subject $i=1,\dots, n$ a potentially right-censored follow-up time $T_i$ and the event indicator $\delta_{i}$, which is 1 if subject $i$ experiences the event and 0 if it is censored. The hazard of an event at time $t$ is modeled by structured additive predictors $\eta_k$, $k \in \{\lambda, \gamma, \alpha, \mu\}$ as 

\begin{equation}
h_{i}(t)= \exp \left\{\eta_i(t)\right\} = \exp\left\{\eta_{\lambda i}(t)+
\eta_{\gamma i}+\eta_{\alpha i}\left(\eta_{\mu i}(t), t\right) \right\}
\label{eq:hazard} 
\end{equation}
with $\eta_{\lambda}$ the predictor for all time-varying survival covariates and effects including the log baseline hazard, $\eta_{\gamma}$ representing time-constant effects of baseline survival covariates, the longitudinal marker $\eta_{\mu}$ and the potentially nonlinear association between the longitudinal marker and the hazard $\eta_\alpha$. Note that by modeling the latter as a function of $\eta_\mu$ and time $t$, a variety of association structures can be specified.\\
In general, the vector of predictors for all subjects is expressed as $\bm{\eta}_k = [\eta_{k1}, \cdots, \eta_{kn}]^{\top}$, $k \in \{\lambda, \gamma, \alpha, \mu, \sigma\}$. In the longitudinal part of the model, the predictor vector is $\bm{\eta}_k(\bt)$, $k \in \{\mu, \sigma\}$ of length $N=\sum_i n_i$ containing entries $\eta_{k i}(t_{ij})$ for all $j=1,\dots, n_i$ per subject $i$, i.e.~corresponding evaluations at all observed time points $\bt=[\bt^{\top}_1, \cdots, \bt^{\top}_n]^{\top}$. In the survival part of the model, the predictor vector $\bm{\eta}_\mu(t)$ is of length $n$ containing one observation per subject at time $t$. This setup in the survival part is analogous to the setup for the other predictors in the survival submodel and additionally, $\bm{\eta}_k(\bct)$ denotes the evaluation of the respective predictor at the vector of follow-up times for all subjects $\bct = [T_1, \dots, T_n]^\top$. 

Each predictor $\eta_{ki}$ with $k \: \in \: \{\lambda, \gamma, \mu, \sigma\}$ is a structured additive predictor $\eta_{ki} =  \sum_{m=1}^{M_k} f_{km}(\bm{\tilde{x}}_{kmi})$ of $M_k$ functions of covariates $\bm{\tilde{x}}_i$. Each function $f_{km}$ depends on one or two covariates, i.e. different subsets $\bm{\tilde{x}}_{kmi}$ of $\bm{\tilde{x}}_i$. For time-varying predictors the functions $\eta_{ki}(t) =  \sum_{m=1}^{M_k} f_{km}(\bm{\tilde{x}}_{kmi}(t), t)$ can also depend on time. By using suitable (e.g.\ spline) basis matrices $\bcx_{km}$ for every term $m$ of predictor $k$ and corresponding penalty $\bcp_{km}$ a variety of effects such as nonlinear, spatial, time-varying or random effects can be modeled under the generic structure

\begin{equation}
\bm{f}_{km} = \bcx_{km} \bm{\beta}_{km} \text{\qquad and \qquad} \bcp_{km}=\frac{1}{\tau_{km}^2} \bm{\beta}_{km}^{\top} \bck_{km} \bm{\beta}_{km}.
\label{eq:penalty}
\end{equation}
Here $\bm{f}_{km}$ denotes the vector of function evaluations stacked over subjects, $\bcx_{km}$ are the design matrices of size $n \times p_{km}$ or $N \times p_{km}$ for the survival and longitudinal submodel, respectively, and  $\bm{\beta}_{km} = [\beta_{km1}, \cdots ,\beta_{kmp_{km}}]^\top$ denotes the coefficient vector of length $p_{km}$. Note that $\bm{x}_{kmi}$ denotes the $i$-th row of the design matrix $\bcx_{km}$ whereas $\bm{\tilde{x}}_{kmi}$ denotes the respective covariate vector. For parametric terms these two often coincide, whereas for spline representations of smooth covariate effects  or random effects terms $\bm{x}_{kmi}$ represents the respective basis evaluation vector of  $ \bm{\tilde{x}}_{kmi}$. For example, random intercepts are modeled using the design matrix $\bcx_{km}$, an $N \times n$ indicator matrix with the $i$th column indicating which longitudinal measurements belong to subject $i$, the coefficient vector $\bm{\beta}_{km} = [ \beta_{km 1}, \cdots , \beta_{km n} ]$ and the penalty matrix $\bck_{km}=\bm{I}_n$, which is an $n \times n$ identity matrix. This penalty ensures $\beta_{km i} \sim N(0, \tau^2_{km} )$ independently.  For the setup of smooth effects using P-splines we refer to the next subsection and details on the setup of the predictors, function evaluations and design matrices for the submodels can be found in the Appendix and in \cite{kohler_flexible_2017}. 

All effects are modeled within a Bayesian framework by specifying appropriate prior distributions for the coefficient vectors, as presented in more detail in Section \ref{sec3}.

\subsection{Flexible associations}
The special focus in the generalization of the hazard in \eqref{eq:hazard}  lies on the flexible specification of the predictor $\eta_{\alpha}$ to incorporate not only time-varying and covariate-dependent associations $\eta_{\alpha i}(t) \cdot \eta_{\mu i}(t)$ as previously presented \citep{kohler_flexible_2017}, which assume an association linear in $\eta_{\mu i}(t)$, but also nonlinear associations between the predicted longitudinal marker and the time-to-event process. 

The general predictor is formulated as $\eta_{\alpha i}( \eta_{\mu i}(t), t) =  f_{\alpha}\left(\eta_{\mu i}(t), \bm{\tilde{x}}_{\alpha i}, t\right)$, that is a function of the potentially smooth time-varying predicted marker trajectories $\eta_{\mu i}(t)$ from \eqref{eq:long},  further covariates $\bm{\tilde{x}}_{\alpha i}$ as well as time $t$. Note that we drop the subscript $m$ whenever this is simpler, as $M_\alpha = 1$. We make use of a suitable basis representation to incorporate this flexible specification into our framework as
\begin{equation}
\label{general}
f_{\alpha}\left(\eta_{\mu i}(t), \bm{\tilde{x}}_{\alpha i}, t\right) =  \left[\bm{g}_1(\eta_{\mu i}(t)) \odot \bm{g}_2(\bm{\tilde{x}}_{\alpha i}, t)\right] \bm{\beta}_{\alpha} = \bm{x}^\top_{\alpha i} \bm{\beta}_{\alpha}
\end{equation} 
with $\odot$ denoting the row tensor product. The row tensor product of a $p \times a$ matrix $\bm{A}$ and a $p \times b$ matrix $\bm{B}$ is defined as the $p \times ab$ matrix $\bm{A} \odot \bm{B}=(\bm{A} \otimes \bm{1}_b^\top ) \cdot (\bm{1}_a^\top \otimes \bm{B})$ with $\cdot$ denoting element-wise multiplication and $\otimes$ the Kronecker product. In this notation $\bm{g}_1(\eta_{\mu i}(t))$ represents the basis vector of the potentially nonlinear effect of the longitudinal predictor $\eta_{\mu i}(t)$ and $\bm{g}_2(\bm{\tilde{x}}_{\alpha i}, t)$ represents the basis vector for the effects of relevant covariates and/or a smooth function of time $t$. The resulting design vector $\bm{x}_{\alpha i}$ and parameter vector $\bm{\beta}_{\alpha}$ are of length $p_\alpha = p_{\alpha 1} \cdot p_{\alpha 2}$.

The standard linear association between the longitudinal predictor and the log-hazard can be formulated as $\bm{g}_1(\eta_{\mu i}(t)) = I(\eta_{\mu i}(t)) = \eta_{\mu i}(t)$, where $I(\cdot)$ denotes the identity, and $\bm{g}_2 \equiv 1$. For nonlinear associations we use P-splines \citep{EilersMarx1996} by specifying a B-spline representation of the longitudinal predictor effect $\bm{g}_1(\eta_{\mu i}(t))= \bcb\left(\eta_{\mu i}(t)\right) = \left[B_{1}\left(\eta_{\mu i}(t)\right), \dots, B_{p_{\alpha 1}}\left(\eta_{\mu i}(t)\right)\right]$. Here, $B_d$ denotes the $d$-th basis function over the observed range of $\eta_{\mu i}(t)$, with $\bm{g}_1(\eta_{\mu i}(t))$ being the corresponding design vector of length $p_{\alpha 1}$ of the spline evaluations at $\eta_{\mu i}(t)$. The corresponding penalty matrix of the effect of $\eta_\mu(t)$ is a zero matrix $\bck_{\alpha 1}= \bf{0}$ for $\bm{g}_1(\eta_{\mu i}(t)) = I(\eta_{\mu i}(t))$ and a P-spline penalty matrix $\bck_{\alpha 1}= \bcd_r^\top\bcd_r$ with the $r$-th difference matrix $\bcd_r$ for $\bm{g}_1(\eta_{\mu i}(t)) = \bcb(\eta_{\mu i}(t))$. 
For simplicity, we denote the function transforming any covariate values $\bm{z}$ into a matrix of evaluations of a spline basis generally as $\bcb(\bm{z})$. This function returns the matrix of respective basis evaluations with number of columns equal to the number of spline basis functions and number of rows equal to the length of $\bm{z}$. 

In order to model simple parametric, nonlinear or time-varying effects, $\bm{g}_2(\bm{\tilde{x}}_{\alpha i}, t)$ can be specified accordingly as a constant, a spline representation of a continuous covariate effect or as spline representation of an effect of time $t$ with appropriate penalty matrix $\bck_{\alpha 2}$. To further illustrate the notation, consider the following effect specifications
\begin{itemize}
\item time-constant, linear association $f_{\alpha}\left(\eta_{\mu i}(t)\right) = \left[I\left(\eta_{\mu i}(t)\right) \odot 1 \right] \beta_{\alpha}$ where $p_{\alpha} = 1 \cdot 1 $,
\item (linearly) covariate-dependent, linear association $f_{\alpha}\left(\eta_{\mu i}(t), \bm{\tilde{x}}_{\alpha i}\right) = \left[I\left(\eta_{\mu i}(t)\right) \odot \bm{\tilde{x}}^\top_{\alpha i} \right] \bm{\beta}_{\alpha}$ where $p_{\alpha} = 1 \cdot p_{\alpha 2} $ with $p_{\alpha 2}$ the length of $\bm{\tilde{x}}_{\alpha i}$,
\item time-varying, linear association $f_{\alpha}\left(\eta_{\mu i}(t), t\right) = \left[I\left(\eta_{\mu i}(t)\right) \odot \bcb(t) \right] \bm{\beta}_{\alpha}$ where $p_{\alpha} = 1 \cdot p_{\alpha 2} $ with $p_{\alpha 2}$ the number of spline basis functions in $\bcb(t)$,
\item time-constant, nonlinear association $f_{\alpha}\left(\eta_{\mu i}(t)\right) = \left[\bcb\left(\eta_{\mu i}(t)\right)^\top \odot 1 \right] \bm{\beta}_{\alpha}$ where $p_{\alpha} = p_{\alpha 1} \cdot 1 $ with $p_{\alpha 1}$ the number of spline basis functions in $\bcb\left(\eta_{\mu i}(t)\right)$,
\item covariate-dependent, nonlinear association $f_{\alpha}\left(\eta_{\mu i}(t), \bm{\tilde{x}}_{\alpha i}\right) = \left[\bcb\left(\eta_{\mu i}(t)\right)^\top \odot \bm{\tilde{x}}^\top_{\alpha i} \right] \bm{\beta}_{\alpha}$ where $p_{\alpha} = p_{\alpha 1} \cdot p_{\alpha 2}$ with $p_{\alpha 1}$ the number of spline basis functions in $\bcb\left(\eta_{\mu i}(t)\right)$ and $p_{\alpha 2}$ the length of $\bm{\tilde{x}}_{\alpha i}$,
\item time-varying, nonlinear association $f_{\alpha}\left(\eta_{\mu i}(t), t\right) = \left[\bcb \left(\eta_{\mu i}(t)\right)^\top \odot \bcb(t) \right] \bm{\beta}_{\alpha}$ where $p_{\alpha} = p_{\alpha 1} \cdot p_{\alpha 2} $ with $p_{\alpha 1}$ and $p_{\alpha 2}$ the number of spline basis functions in $\bcb\left(\eta_{\mu i}(t)\right)$ and $\bcb(t)$, respectively. 
\end{itemize}

For both, time-varying effects and nonlinear associations, Bayesian P-Splines \citep{lang_bayesian_2004-1} are employed where smoothing is induced by appropriate prior specification. In more detail the difference penalties are replaced by their stochastic analogues, i.e. random walks. The full penalty $\bcp_{\alpha}$ allows for different amounts of smoothing across both $\eta_{\mu i}(t)$ and the covariate or time effects by using an anisotropic smooth with
\begin{equation}
\bcp_{\alpha} = \bm{\beta}_{\alpha}^\top \left( \frac{1}{\tau_{\alpha 1}^2} \bck_{\alpha 1} \otimes \bm{I}_{p_{\alpha 2}} + \frac{1}{\tau_{\alpha 2}^2} \bm{I}_{p_{\alpha 1}} \otimes \bck_{\alpha 2} \right) \bm{\beta}_{\alpha} = \bm{\beta}_{\alpha}^\top \left( \frac{1}{\tau_{\alpha 1}^2} \bm{\tilde{K}}_{\alpha 1} + \frac{1}{\tau_{\alpha 2}^2}  \bm{\tilde{K}}_{\alpha 2} \right) \bm{\beta}_{\alpha},
\label{eq:aniso_pen}
\end{equation}
where $\bm{I}_a$ is an $a \times a$ identity matrix. Within the \proglang{R} package \pkg{bamlss} currently all above mentioned linear associations as well as constant and group-specific nonlinear associations are implemented. Further nonlinear associations are under construction.

\subsection{Identifiability}
\label{identifiability}
Given the additive structure of the model and the fact that all model parts always contain an intercept in our construction,  constraints on certain predictors are necessary to obtain an identifiable model. The general constraint for all nonlinear terms in the model is a sum-to-zero constraint over all $n$ or $N$ observations for predictors in the survival and longitudinal submodel, respectively, e.g.\ $\sum_i f_{\lambda m}(T_i)=0$ or $\sum_i f_{\gamma m}(\bm{\tilde{x}}_{\gamma m i})=0$. 
These constraints are implemented for B-splines by transforming the $n \times p_{km}$ basis matrix $\bcx_{km}$ into an $ n \times (p_{km}-1)$ matrix $\bm{\dot{X}}_{km}$ for which it holds that $\bm{\dot{X}}_{km} \bm{1}_{p_{km}-1}= \bm{0}$ as shown in \citet[][chapter 1.8]{wood_generalized_2006}, and adjusting the penalty accordingly. For tensor product smooth terms the constraint is achieved by transforming the marginal basis matrices and the corresponding marginal penalties. For example, the constraint for functional random intercepts in $\eta_{\mu i}(t)$ is achieved by transforming the marginal basis matrix of the smooth effect of time and the corresponding marginal penalty as above. 
In the case of a nonlinear specification of $\eta_{\alpha i}(\eta_{\mu i}(t))$ the marginal basis in $\bm{g}_1$ is constrained slightly different. As the predictor $\eta_{\mu i}(t)$  and therefore also its spline basis evaluation is estimated within the model, we choose a constraint based on the observed marker. In more detail, we constrain the term to sum to zero on a fixed grid $\bm{y}^\ast$ from the 2.5th to the 97.5th quantile of the observed longitudinal response, i.e. $\bm{1}^\top \bm{\eta}_\alpha(\bm{y}^\ast)  = 0$ with $\bm{1}$ a vector of ones. For nonlinear effects per factor level $g$ the same constraint is enforced for every level $g$ and one intercept per factor level except the reference level is included in the model.

\section{Estimation}
\label{sec3}

We estimate the model in a Bayesian framework using a Newton-Raphson procedure and a derivative-based Markov chain Monte Carlo (MCMC) algorithm to estimate the mode and the mean of the posterior distribution of the vector $\bm{\theta}$ of all parameters, respectively. 

Assuming conditional independence of the survival outcomes $[T_i, \delta_i]$ and the longitudinal outcomes $\bm{y}_i$, given the parameters $\bm{\theta}$, the posterior of the full model is
\begin{equation}
\begin{split}
p(\bm{\theta} | \bct, \bm{\delta}, \bm{y}) \propto L^{\text{long}}\left[\bm{\theta} | \bm{y}\right] \cdot L^{\text{surv}}\left[\bm{\theta} | \bct, \bm{\delta}\right] \prod_{k \in \{\lambda, \gamma, \alpha, \mu, \sigma\} } \prod_{m=1}^{M_k} \left[p(\bm{\beta}_{km}|\bm{\tau}_{km}^2)  p(\bm{\tau}_{km}^2)\right],
\end{split}
\end{equation}
with the likelihood of the longitudinal submodel $L^{\text{long}}$  \eqref{eq:long} and the survival submodel $L^{\text{surv}}$  \eqref{eq:hazard}, and the response vectors $\bm{y} = [\bm{y}_1^\top, \cdots, \bm{y}_n^\top]^\top$ and $\bm{\delta}=[\delta_1, \cdots ,\delta_n]^{\top}$. Further, $p(\bm{\beta}_{km}|\bm{\tau}_{km}^2)$ and $p(\bm{\tau}_{km}^2)$ denote the priors of the vectors of regression parameters and variance parameters for each term $m$ and predictor $k$. Note that for anisotropic smooths, multiple variance parameters are used resulting in the vector $\bm{\tau}^2_{km}$. 

\subsection{Likelihood}
The log-likelihood of the longitudinal part is 
\begin{equation}
\ell^{\text{long}}\left[\bm{\theta} | \bm{y} \right] = -\frac{N}{2}\log(2\pi)- \bm{1}_{N}^{\top} \bm{\eta}_{\sigma}\left(\bt\right) -\frac{1}{2}(\bm{y}-\bm{\eta}_{\mu}\left(\bt\right))^{\top} \bm{R}^{-1}(\bm{y}-\bm{\eta}_{\mu}\left(\bt\right))
\label{eq:l_long}
\end{equation}
where $\bm{R}=\text{blockdiag} (\bm{R}_{1}, \cdots, \bm{R}_{n})$. $\bm{R}$ simplifies to a diagonal matrix, as we assume $\bm{R}_{i}=\diag (\exp[\eta_{\sigma i}(t_{i1})]^2, \cdots, \exp[\eta_{\sigma i}(t_{in_i})]^2)$. 

The log-likelihood of the survival part of the model is
\begin{equation}
\ell^{\text{surv}}\left[\bm{\theta} | \bct, \bm{\delta}\right] =  \bm{\delta}^{\top}\bm{\eta}(\bct)-\bm{1}_{n}^{\top}\bm{\Lambda}\left(\bct\right)
\end{equation}
where $\bm{\Lambda}(\bct)=[\Lambda_{1}(T_{1}),\cdots,\Lambda_{n}(T_{n})]^{\top}$ denotes the vector of cumulative hazard rates with  $\Lambda_{i}(T_{i})=$ \\
$\exp(\eta_{\gamma i})\int_{0}^{T_{i}}\exp[\eta_{\lambda i}(u)+\eta_{\alpha i}(\eta_{\mu i}(u), u)]du$.

\subsection{Priors}

In our setup different terms, such as smooth, time-varying or random effects, are specified by the choice of corresponding design matrices and priors. For linear or parametric terms we use vague normal priors  on the vectors of the regression coefficients, e.g.\ $\bm{\beta}_{km} \sim N(\bm{0}, 1000^2 \bm{I})$, to approximate a precision matrix $\bck_{km}=\bm{0}$. Multivariate normal priors 
\begin{equation}
p(\bm{\beta}_{km}|\tau_{km}^2) \propto \left(\frac{1}{\tau_{km}^2}\right)^{\frac{\operatorname{rank}\left(\bck_{km}\right)}{2}} \exp \left(-\frac{1}{2\tau_{km}^2} \bm{\beta}_{km}^\top \bck_{km} \bm{\beta}_{km} \right)
\end{equation}
are used to regularize smooth and random effect terms with precision matrix $\bck_{km}$ as specified in the penalty \eqref{eq:penalty}. 
For anisotropic smooths as in the flexible association $\eta_\alpha$ in   \eqref{eq:aniso_pen}, when multiple variance parameters are involved, e.g.\ $\bm{\tau}_{\alpha}^2 = (\tau_{\alpha 1}^2, \tau_{\alpha 2}^2)$, we use the prior
\begin{equation}
p(\bm{\beta}_{km}|\bm{\tau}_{km}^2) \propto \left|\frac{1}{\tau_{km1}^2}    \bm{\tilde{K}}_{km1}   + \frac{1}{\tau_{km2}^2}  \bm{\tilde{K}}_{km2}  \right|^{\frac{1}{2}}
\exp \left( -\frac{1}{2} \boldsymbol{\beta}_{km}^\top \left[\frac{1}{\tau_{km1}^2}    \bm{\tilde{K}}_{km1}  + \frac{1}{\tau_{km2}^2}  \bm{\tilde{K}}_{km2}  \right] \boldsymbol{\beta}_{km}  \right).
\label{eq:aniso_prior}
\end{equation}

As priors for the variance parameters $\tau^2_{km}$, which control the trade-off between flexibility and smoothness in the nonlinear modeling of effects, we use independent inverse Gamma hyperpriors $\tau_{km}^2 \sim IG(0.001, 0.001)$ to obtain an inverse Gamma full conditional (component-wise in the case of variance vectors $\bm{\tau}_{km}^2)$. Further priors for the variance parameters, such as half-Cauchy, are possible.

\subsection{Posterior Mode and Posterior Mean}
To obtain starting values for the posterior mean estimation and to gain a quick model assessment we estimate the mode of the posterior using a Newton-Raphson procedure. In more detail, we maximize the log-posterior by updating 
blockwise each term $m$ of predictor $k$ in each iteration $l$ as
\begin{equation}
\bm{\beta}_{km}^{[l+1]}=\bm{\beta}^{[l]}_{km}- \nu^{[l]}_{km} \bm{H}\left(\bm{\beta}_{km}^{[l]}\right)^{-1}\bm{s}\left(\bm{\beta}_{km}^{[l]}\right)
\end{equation}
with steplength $\nu^{[l]}_{km}$, the score vector $\bm{s}(\bm{\beta}_{km} )$ and the Hessian $\bm{H}(\bm{\beta}_{km})$. In each updating step we optimize the steplength $\nu_{km}^{[l]}$ over $(0, 1]$ to maximize the log-posterior and the variance parameters to minimize the corrected AIC \citep[AICc,][]{hurvich_smoothing_1998}. The block-wise score vectors and Hessians can be found in the Appendix. For quick approximate inference we derive credibility intervals from $N(\hat{\bm{\beta}}_{km}, [-\bm{H}(\hat{\bm{\beta}}_{km}) ]^{-1} )$ assuming an approximately normal posterior distribution for the coefficients $\bm{\beta}_{km}$. Note however, that as these credibility intervals do not take into account the optimization of the variance parameters, they tend to underestimate the variability and posterior mean sampling should be used for exact inference.

The focus of our model estimation lies on the derivative-based Metropolis-Hastings posterior mean sampling. We construct approximate full conditionals $\pi(\bm{\beta}_{km}| \cdot)$ based on a second order Taylor expansion of the log-posterior centered at the last state $\beta_{km}^{[l]}$ as shown in \cite{umlauf_bamlss_2017}. 
This approximate full conditional results in a multivariate normal proposal density with the precision matrix $(\bm{\Sigma}^{[l]}_{km})^{-1} = -\bm{H}(\bm{\beta}^{[l]}_{km})$ and the mean $\bm{\mu}^{[l]}_{km} = \bm{\beta}^{[l]}_{km} - \bm{H}(\bm{\beta}^{[l]}_{km})^{-1} \bm{s}(\bm{\beta}_{km}^{[l]})$. We draw a candidate $\bm{\beta}_{km}^\ast$ from the proposal density $q(\bm{\beta}^\ast_{km} | \bm{\beta}^{[l]}_{km}) = N(\bm{\mu}_{km}^{[l]},\bm{\Sigma}^{[l]}_{km})$ in each iteration $l$ of the Metropolis-Hastings sampler for updating block $km$. Despite being computationally demanding, drawing candidates from a close derivative-based approximation of the full conditional results in high acceptance rates and good mixing as we approximate a Gibbs sampler. Samples for the variance parameters $\tau^2_{km}$ are either obtained via Gibbs sampling, if inverse Gamma hyperpriors are used and the full conditionals $\pi(\tau^2_{km}|\cdot)$ in consequence follow an inverse Gamma distribution, or via slice sampling when no simple closed-form full conditional can be obtained. This is the case in the sampling of variance parameters for anisotropic smooths or when other hyperpriors than the inverse Gamma are used. We suggest to use DIC for model selection.

\section{Simulation}
\label{sec4}

The performance of the presented framework is tested in extensive simulations of which a subset of the results is shown in the following. Three main questions motivated the simulations: First, we aim to assess how well the flexible joint model can estimate truly linear associations, also in comparison to established implementations as in the \proglang{R} package \pkg{JMbayes}. Second, we explore how well the model can capture truly nonlinear associations and assess the extent of the bias if the nonlinear association is falsely modeled as linear in the log-hazard in \pkg{JMbayes}. Third, the performance of fitting a nonlinear effect per subgroup is assessed. As previous work has shown a strong dependence of the estimation precision on the number of subjects, we test data sets of two different sizes in all three simulation settings. 

\subsection{Simulation design}
We simulate data according to \eqref{eq:long} and \eqref{eq:hazard} where we use in setting 1 the linear association  $\eta_{\alpha i}(\eta_{\mu i}(t)) = 1 \cdot \eta_{\mu i}(t) $ between the longitudinal marker and the log-hazard, in setting 2 the nonlinear association $\eta_{\alpha i}(\eta_{\mu i}(t)) = -0.1(\eta_{\mu i}(t) + 3)^2 + \eta_{\mu i}(t) + 1.8$ and in setting 3 a group-specific nonlinear association $\eta_{\alpha i}(\eta_{\mu i}(t), g_i = 1) = -0.1(\eta_{\mu i}(t) + 3)^2 + \eta_{\mu i}(t) + 1.8$ and $\eta_{\alpha i}(\eta_{\mu i}(t), g_i = 0) = 0.1(\eta_{\mu i}(t)-3)^2 + 0.75\eta_{\mu i}(t) - 0.8$. In all settings we generate $Q = 200$ data sets with $n = 300, 600$, respectively, to assess the influence of sample size on the precision of the estimates.

In more detail we generate longitudinal marker values $\eta_{\mu i}(t) = \sum_{m = 1}^{5} f_{\mu m}(\bm{\tilde{x}}_{\mu m i}, t)$ at a fixed grid of timepoints $t^\ast = 1, \dots, 120$ with the time effect $ f_{\mu 1}\left(t\right) = 0.1(t+2)\exp(-0.075t)$, random intercepts $f_{\mu 2}\left(i\right) = r_{i}$ where $r_{i} \sim N(0, 0.25)$, functional random intercepts (i.e. smooth subject-specific trajectories) $f_{\mu 3}\left(t, i\right)= \bcx_{\mu 3} \bm{\beta}_{\mu 3} =  (\bcx_{\mu 3 s} \odot \bcx_{\mu 3 t}) \bm{\beta}_{\mu 3}$ where $\bcx_{\mu 3 s}$ and $\bcx_{\mu 3 t}$ are the basis representations of a random intercept  and a spline over $t$, respectively, as well as a global intercept $f_{\mu 4}(\bm{x}_{\mu i}) = 0.5$ and covariate effect $ f_{\mu 5}(\bm{x}_{\mu i}) = 0.6 \sin(x_{2i})$ with $x_{2i} \sim U(-3, 3)$. The functional random intercepts are simulated using P-Splines based on cubic B-splines where the true vector of spline-coefficients with 4 basis functions per subject is drawn from $\bm{\beta}_{\mu 3} \sim N(\bm{0},[(1/ \tau_{\mu 3 s}^2) \bm{\tilde{K}}_{\mu 3 s} + (1/ \tau_{\mu 3 t}^2)  \bm{\tilde{K}}_{\mu 3 t}]^{-1})$ where  $\bm{\tilde{K}}_{\mu 3 s} = \bck_{\mu 3 s} \otimes \bm{I}_4$ with  $ \bck_{\mu 3 s}=\bm{I}_n$ as the penalty matrix for the random effect and $\bm{\tilde{K}}_t = \bm{I}_n \otimes \bck_{\mu 3 t} $ with $\bck_{\mu 3 t}$ as an appropriate penalty matrix for the smooth effect of time with $\bck_t = \bcd_2^\top \bcd_2$, $\tau_s^2=1$ and $\tau_t^2=0.2$. Similar to \eqref{eq:aniso_pen} the two marginal penalties enter into a Kronecker sum penalty.  

We calculate the hazard $h_i(t)$ for every subject using $\eta_{\lambda}(t)= 1.4\log((t+10)/1000)$,  $\eta_{\gamma i}=0.3x_{1i}$, with $x_{1i} \sim U(-3, 3)$ and $\eta_{\alpha}$ as described above. Survival times for every subject are derived using survival probabilities obtained by numerical integration as described in \citet{bender_generating_2005} and \citet{crowther_simulating_2013} and censored at $t = 120$. We additionally censor all survival times uniformly using $U(0, 1.5 \cdot 120)$. In order to mimic the irregular measurement times we randomly delete 90\% of the generated longitudinal measurements resulting in a median of 6 measurements per subject (interquartile range (IQR): 3, 10) for every setting. Finally we obtain longitudinal observations $y_{ij}$ from $\eta_{\mu i}(t)$ by adding independent errors $\epsilon_{ij} \sim N(0, 0.3^2)$ for each $t_{ij}$ in $\bt$. As the estimation showed stability issues in the most complex model in setting 3 for small samples we fit setting 3 also leaving more longitudinal observations by deleting only 80\% of the simulated observations resulting in a median of 12 measurements per subject (IQR: 6, 18).

We fit the 1600 generated data sets ($(3+1)$ settings $\times$ 2 sample sizes $\times$ 200 replications) with our model implementation in \pkg{bamlss} where setting 3 is simulated using a median of 6 and 12 measurements per subject. Additionally we compare our results in setting 1 and 2 with the linear estimation in \pkg{JMbayes}. 
For \pkg{bamlss} we estimate the longitudinal trajectories using P-splines \citep{EilersMarx1996} with cubic B-Splines, a second order difference penalty and 10 knots (2 internal knots) for the overall mean and the individual trajectories resulting in $5 \cdot n$ basis functions. The association $\eta_{\alpha i}(\eta_{\mu i}(t))$ is also modeled using P-Splines with 5 basis functions after imposing the constraint in setting 1 and 2, and for each of both groups in setting 3. In a few cases the posterior mode estimation led to extreme predictions in $\eta_{\mu i}(t)$ for single subjects. In these cases we reduced the number of coefficients for $\eta_{\alpha i}(\eta_{\mu i}(t))$ by 2 to stabilize the estimation. 
This occurred 2 and 4 times in setting 1, for small and large data sets, respectively, 3 and 2 times in setting 2, and 5 and 1 times in setting 3 with a median of 6 observations per person as well as 2 times each with a median of 12 observations per person. 
Further, the baseline hazard $\eta_\lambda$ is estimated using P-Splines with 9 resulting basis functions. For setting 3 we allow the nonlinear association to vary between the two subgroups $\eta_{\alpha i}(\eta_{\mu i}(t), g_i)$. For comparison we also fit the data sets assuming a linear association with the log-hazard using \pkg{JMbayes} in setting 1 and 2 and try to achieve otherwise comparable models by modeling the nonlinear effects in the longitudinal submodel by the available unpenalized B-splines and the baseline hazard by P-splines. The number of knots were assessed in preliminary simulations to minimize the AIC resulting in 3 basis functions per subject with diagonal covariance matrix of the random effects for  $n = 300$ and 4 basis functions per subject for $n = 600$. For the posterior mean estimation we sample for 13000 iterations, discard 3000 samples as burnin and keep 5000 samples per model after thinning. 

In every estimated model we calculate mean-squared error (MSE), bias, and frequentist coverage of the 95\% credibility interval both averaged over all time points and averaged per time point. For the predictors in the longitudinal model, i.e. $k \in \{\mu, \sigma\}$, the average MSE in each sample $q$ is $MSE^q_{k} = \frac{1}{N} \sum_{i=1}^{n} \sum_{j=1}^{n_i} [\eta^{q}_{k i}(t_{ij}) - \hat{\eta}^{q}_{k i}(t_{ij})]^2$ with the estimate $\hat{\eta}_{ki}$, and the MSE per time point is $MSE^q_k(t) =  \frac{1}{n} \sum_{i=1}^{n} [\eta^{q}_{k i}(t) - \hat{\eta}^{q}_{k i}(t)]^2$ for all $t$ in $t^\ast$. For the survival predictors $\eta_\gamma$ and $\eta_\lambda$, the average MSE is $MSE^q_{k} = \frac{1}{n} \sum_{i=1}^{n} [\eta^{q}_{k i}(T_{i}) - \hat{\eta}^{q}_{k i}(T_{i})]^2$ using evaluations at the subject's event times for $\eta_\lambda$ and for the time-constant $\eta_\gamma$. For $\eta_{\lambda}$ the error is additionally evaluated at the fixed grid of time points $t^\ast$ as above. For the potentially nonlinear association $\eta_{\alpha i}(\eta_{\mu i}(t))$ a variety of different evaluations are possible. As the association is a survival predictor we compute the average error as $MSE^q_{\alpha} = \frac{1}{n} \sum_{i=1}^{n} [\eta^{q}_{\alpha i}(\eta_{\mu i}(T_{i})) - \hat{\eta}^{q}_{\alpha i}(\eta_{\mu i}(T_{i}))]^2$. To assess the performance over the full range of the marker values and to assess deviations from a linear fit we also compute $MSE^q_{\alpha}(\eta^\ast_{\mu}) =  [\eta^{q}_{\alpha}(\eta^\ast_{\mu}) - \hat{\eta}^{q}_{\alpha}(\eta^\ast_{\mu})]^2$  where $\eta^\ast_{\mu}$ is from a fixed grid from -0.5 to 2 in 120 steps. This fixed grid was chosen as the maximum range of true values $\eta_\mu$ that were simulated in all settings. For setting 3, this measure is computed per group and then averaged over groups. All these error measures are then averaged over all $Q$ samples per setting. Additionally we compute a point estimate of the average slope of the association as the averaged first derivative $\frac{1}{n}\sum_{i = 1}^n\eta_{\alpha i}^{\prime}(\eta_{\mu i}(T_i))$ of the estimated association in setting 1.

\subsection{Simulation results}

In setting 1 \pkg{bamlss} allows for an unbiased modeling of the linear association with satisfactory frequentist coverage of the credibility bands. All survival predictors show systematically less estimation error when more information is available as for $n =600$ (cf. Table \ref{tbl:sim1}). Only the predictor $\eta_\sigma$ has a coverage clearly below 0.95. However, as inference for this predictor is rarely of interest, this deviation is not deemed problematic. \pkg{JMbayes} achieves similar performance in setting 1 for most predictors, however the coverage for $\eta_\lambda$ and also $\eta_\alpha$ in the smaller data setting is below the nominal 0.95, especially for $n = 300$.  

\begin{table}[htb]
\centering
\begin{threeparttable}
\caption{Posterior mean estimation results from \pkg{bamlss} and \pkg{JMbayes} from setting 1 (linear $\eta_{\alpha}$) for small and large data sets.}
\label{tbl:sim1}
\begin{tabular}{llcccccc}
\hline
&  & \multicolumn{2}{c}{MSE} & \multicolumn{2}{c}{bias} & \multicolumn{2}{c}{coverage} \\ 
& & $n = 300$ & $n =600$ & $n = 300$ &  $n =600$ & $n = 300$ & \multicolumn{1}{c}{ $n =600$} \\ 
\hline
$\eta_\alpha$ & \pkg{bamlss} & $0.025$ & $0.016$ & $-0.005$ & $-0.006$ & $0.976$ & $0.958$ \\
              & \pkg{JMbayes}& $0.016$ & $0.007$ & $\phantom{-}0.002$ & $\phantom{-}0.000$ & $0.930$ & $0.944$ \\
$\eta_\gamma$ & \pkg{bamlss} & $0.020$ & $0.010$ & $-0.003$ & $\phantom{-}0.018$ & $0.953$ & $0.951$ \\
              & \pkg{JMbayes}& $0.021$ & $0.010$ & $-0.016$ & $\phantom{-}0.015$ & $0.950$ & $0.954$ \\
$\eta_\lambda$& \pkg{bamlss} & $0.042$ & $0.024$ & $-0.000$ & $\phantom{-}0.000$ & $0.948$ & $0.951$ \\
              & \pkg{JMbayes}& $0.043$ & $0.024$ & $\phantom{-}0.000$ & $\phantom{-}0.000$ & $0.915$ & $0.933$ \\
$\eta_\mu$    & \pkg{bamlss} & $0.031$ & $0.031$ & $-0.001$ & $\phantom{-}0.000$ & $0.946$ & $0.946$ \\
              & \pkg{JMbayes}& $0.039$ & $0.030$ & $-0.000$ & $\phantom{-}0.010$ & $\ast$ & $\ast$ \\
$\eta_\sigma$ & \pkg{bamlss} & $0.001$ & $0.001$ & $\phantom{-}0.013$ & $\phantom{-}0.014$ & $0.898$ & $0.859$ \\
              & \pkg{JMbayes}& $0.009$ & $0.000$ & $\phantom{-}0.093$ & $\phantom{-}0.008$ & $\ast$ & $\ast$ \\
\hline 
\end{tabular}
    \begin{tablenotes}
      \small
      \item $\ast$ No credibility intervals and thus no coverage could be calculated for these predictors. 
      \item Results are based on 186 estimates for $n = 300$ and 198 estimates for $n = 600$ .
    \end{tablenotes}
\end{threeparttable}
\end{table}

\begin{figure}[htb]
\centering
\includegraphics[width=0.8\textwidth]{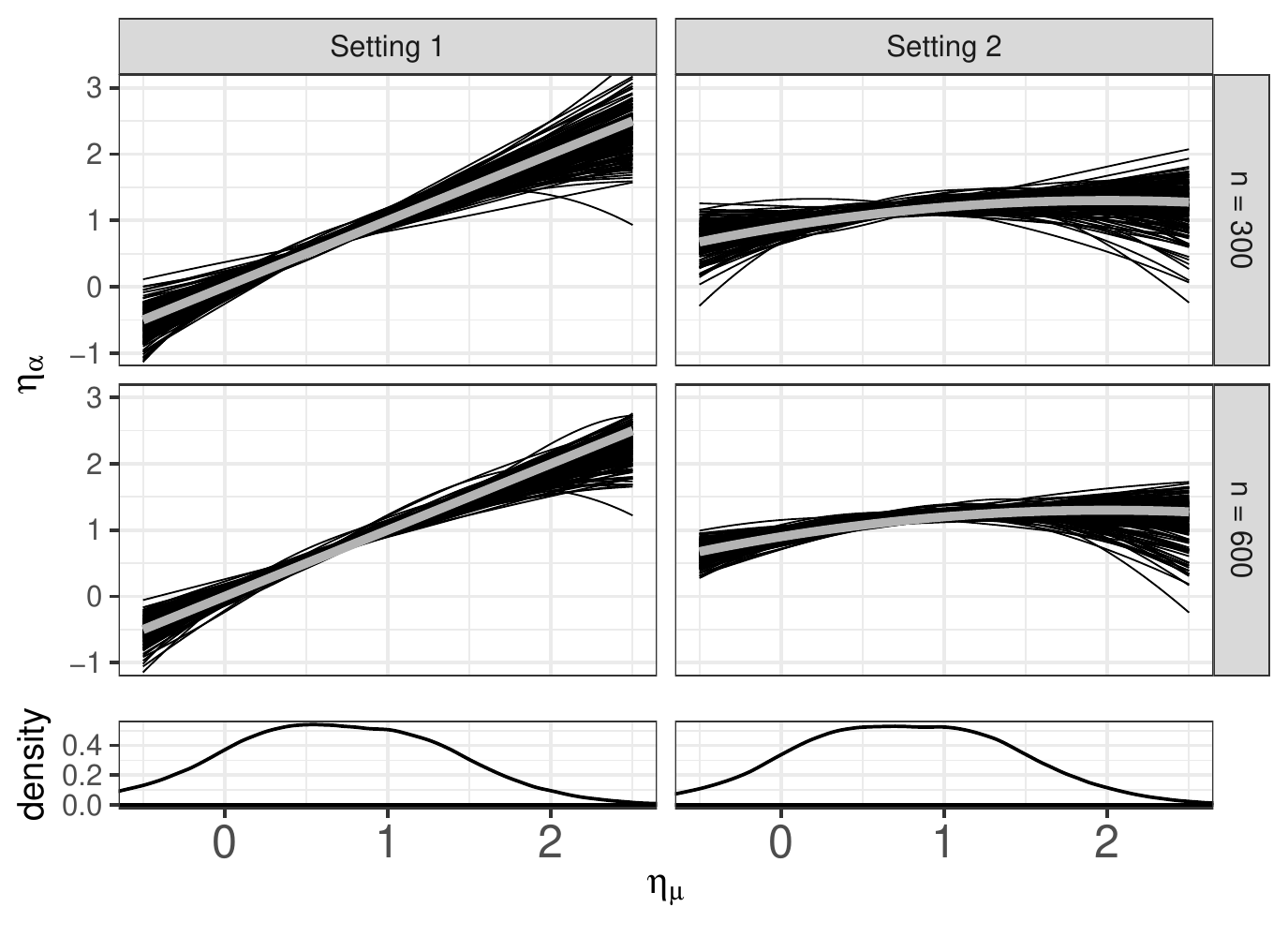}
\caption{True (grey) and estimated (black) predictors from posterior mean estimation of setting 1 (true linear) and setting 2 (true nonlinear) for $n = 300$ and $n = 600$ as well as respective densities of true $\eta_\mu$. Displayed effects are subject to centering constraints as explained in \ref{identifiability}}
\label{fig:sim1_alpha}
\end{figure}

The nonlinearly estimated association from \pkg{bamlss} shows a higher MSE than the linear estimation from \pkg{JMbayes}. As previous simulations \citep{kohler_flexible_2017} have shown the good estimation performance of our implementation, comparable or better than for \pkg{JMbayes},
this difference in MSE is likely caused by the more flexible model specification in \pkg{bamlss}. Overall the association is captured well in our implementation with a mean over all calculated average slopes of 0.99 [average 2.5\% and 97.5\% quantile of the posterior: 0.68; 1.32] for  $n = 300$ and 0.96 [0.75; 1.19] for  $n = 600$. The estimates  show less variability when more data is available, both when more subjects are observed and in areas where more observations of $\eta_\mu$ are measured (see left panel of Figure \ref{fig:sim1_alpha}).  These results are highly similar to the respective linear estimates of \pkg{JMbayes} of 1.02 [0.74; 1.31] and 0.99 [0.79; 1.19], respectively. Note also that the difference of the average quantiles is not much larger for \pkg{bamlss} despite a more flexible model formulation.

The estimation of the nonlinear model in \pkg{bamlss} shows some stability issues when less data is available such that initially 10\% of the estimations for  $n = 300$ and 4\% of the estimations for  $n = 600$ failed as they got stuck in areas of the parameter space where the Hessian for $\bm{\beta}_\mu$ was no longer negative definite.
When restarting the algorithm in such cases with a different seed, these error rates decreased to 7\% and 1\%, respectively. 
Due to the flexibility in the model, especially in the random functional intercepts, the estimation of \pkg{bamlss} takes on average 3.6 and 7.3 hours for $n = 300$ and $n = 600$, respectively, compared to 4 and 7 minutes for \pkg{JMbayes} on a single core of a 2.6GHz Intel Xeon Processor E5-2650. This computation time can be reduced by using more than one core in the MCMC sampling in \pkg{bamlss} as implemented in the package for Linux systems.

A similar overall pattern is seen in setting 2 (cf. Table \ref{tbl:sim2}) for the estimation of \pkg{bamlss}: All estimates of the survival submodel are better with more data, and the coverage is satisfactory except for $\eta_\sigma$. The nonlinearity is captured in the estimation, as shown in Figure \ref{fig:sim1_alpha}, although there is more uncertainty for very high and very low values of $\eta_\mu$, where few observations are available. For the estimation in \pkg{JMbayes}, assuming linearity, the point estimates for the association are good, at least in this setting with small curvature of the association,  however the coverage is very low under this misspecification with $0.70$ and $0.66$ for $n = 300$ and $n = 600$. 

Again some stability issues emerge for \pkg{bamlss} where initially 15.5\%[5\%] of the estimations fail in the smaller[larger] data setting, which was reduced to 7.5\%[3\%] by restarting the estimation with a different seed.  Similarly to setting 1 the estimation takes on average 3.9 and 7.2 hours for $n = 300$ and $n = 600$, respectively.

\begin{table}[htb]
\centering
\begin{threeparttable}
\caption{Posterior mean simulation results from \pkg{bamlss} and results from \pkg{JMbayes} from setting 2 (nonlinear $\eta_{\alpha}$) for small and large data sets.}
\label{tbl:sim2}
\begin{tabular}{llcccccc}
\hline
& & \multicolumn{2}{c}{MSE} & \multicolumn{2}{c}{bias} & \multicolumn{2}{c}{coverage} \\ 
& & $n = 300$ & $n =600$ & $n = 300$ &  $n =600$ & $n = 300$ & \multicolumn{1}{c}{ $n =600$} \\ 
\hline
$\eta_\alpha$& \pkg{bamlss}  & $0.018$ & $0.011$ & $\phantom{-}0.004$ & $\phantom{-}0.001$ & $0.963$ & $0.963$ \\
             & \pkg{JMbayes} & $0.016$ & $0.010$ & $\phantom{-}0.008$ & $\phantom{-}0.007$ & $0.702$ & $0.656$ \\
$\eta_\gamma$& \pkg{bamlss}  & $0.017$ & $0.009$ & $-0.022$ & $-0.017$ & $0.941$ & $0.955$ \\
             & \pkg{JMbayes} & $0.014$ & $0.007$ & $-0.006$ & $-0.002$ & $0.949$ & $0.955$ \\
$\eta_\lambda$& \pkg{bamlss} & $0.037$ & $0.037$ & $\phantom{-}0.000$ & $\phantom{-}0.000$ & $0.943$ & $0.944$ \\
             & \pkg{JMbayes} & $0.031$ & $0.020$ & $\phantom{-}0.000$ & $\phantom{-}0.000$ & $0.914$ & $0.922$ \\
$\eta_\mu$   & \pkg{bamlss}  & $0.032$ & $0.032$ & $-0.001$ & $\phantom{-}0.000$ & $0.947$ & $0.947$ \\
             & \pkg{JMbayes} & $0.039$ & $0.031$ & $-0.006$ & $\phantom{-}0.003$ & $\ast$ & $\ast$ \\
$\eta_\sigma$& \pkg{bamlss}  & $0.002$ & $0.001$ & $\phantom{-}0.014$ & $\phantom{-}0.012$ & $0.914$ & $0.897$ \\
             & \pkg{JMbayes} & $0.008$ & $0.000$ & $\phantom{-}0.085$ & $\phantom{-}0.007$ & $\ast$ & $\ast$ \\
\hline 
\end{tabular}
    \begin{tablenotes}
      \small
      \item $\ast$ No credibility intervals and thus no coverage could be calculated for these predictors. 
      \item Results are based on 185 estimates for $n = 300$  and 194 estimates for $n = 600$ .
    \end{tablenotes}
\end{threeparttable}
\end{table}

\begin{figure}[htb]
\begin{center}
\includegraphics[width=1\textwidth]{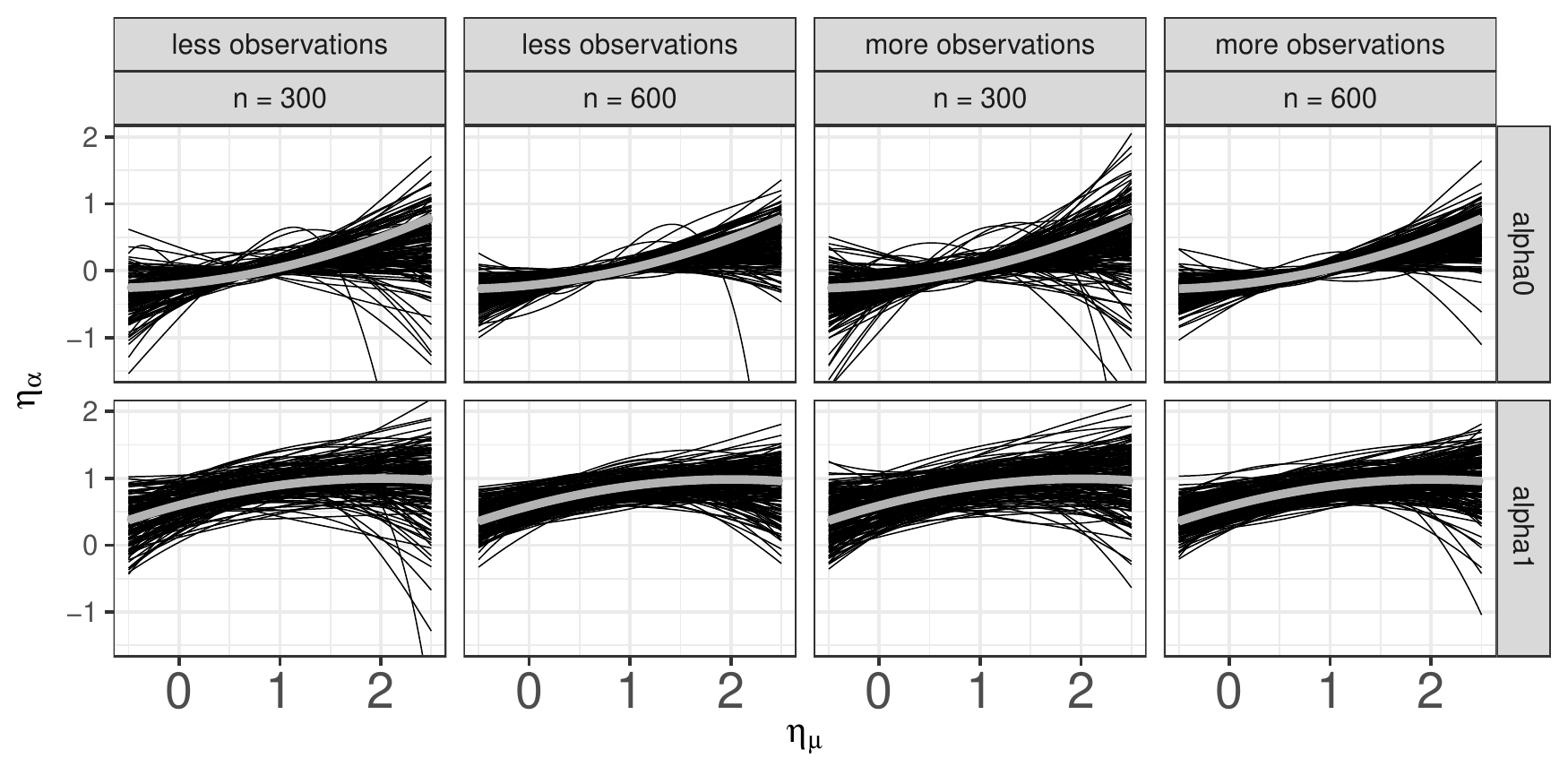}
\caption{True (grey) and estimated (black) predictors from posterior mean estimates of $\eta_{\alpha i}(\eta_{\mu i}(t), g_i)$ in setting 3 for $g_i = 0$ (alpha0) and $g_i = 1$ (alpha1), $n = 300$ and $n = 600$ as well as for a median of 6 longitudinal observations per subject (less observations) and 12 observations per subject (more observations); displayed effects are subject to centering constraints as explained in Section \ref{identifiability} and alpha1 additionally includes the group intercept relative to the reference group $g_i = 0$.} 
\label{fig:sim2_alpha}
\end{center}
\end{figure}

In the most complex model in setting 3 where the association is nonlinear and group-specific, $\eta_{\alpha i}(\eta_{\mu i}(t), g_i)$, the estimation of this association is less precise and more variable than in setting 2 (cf. Table \ref{tbl:sim3} as well as Figure \ref{fig:sim2_alpha}). The precision of the association estimate is generally higher for more subjects, with more longitudinal observations per subject and especially in the areas where $\eta_\mu$ is more densely observed. As in the previous simulations, less information about $\eta_\mu$ was available for the lower and higher values. The estimates are mainly unbiased with only $\eta_\gamma$ showing a  showing a small negative bias for $n = 300$, which is due to an underestimation by on average -0.074 of the global intercept -5.143 in the survival submodel. Further, the credibility intervals show a satisfactory coverage, except for $\eta_\sigma$.

The most complex estimation of a group-specific nonlinear association also suffers most from stability issues such that 43.5\% of the estimations in setting 3 for $n = 300$ as well as 18.5\% of those for $n = 600$ fail for a median of 6 longitudinal observations. These numbers reduced to 30\% and 13\% after restarting the algorithm with a different seed. Included in these problematic estimations are also 2 and 1 estimations, respectively, in which a low acceptance rate (< 30\%) in $\eta_\mu$ indicated sampling issues. In comparison, more observations per subject result in error rates of only 13\% and 4.5\%, which reduced to 5\% and 1\% after restarting with a different seed. Simulations took on average 4.4 and 8.4 hours for $n = 300$ and $n = 600$, respectively, for a median of 6 observations per subject and 4.6 and 9 hours for a median of 12 observations.

\begin{table}[htb]
\begin{threeparttable}
\begin{center}
\caption{Posterior mean simulation results from \pkg{bamlss} for setting 3 (nonlinear, group-specific $\eta_{\alpha}$) using a median of 6 observations per subject or 12 observations per subject.}
\label{tbl:sim3}
\begin{tabular}{llcccccc}
\hline
& & \multicolumn{2}{c}{MSE} & \multicolumn{2}{c}{bias} & \multicolumn{2}{c}{coverage} \\ 
& median $n_i$ & $n = 300$ & $n =600$ & $n = 300$ &  $n =600$ & $n = 300$ & \multicolumn{1}{c}{ $n =600$} \\ 
\hline
$\eta_\alpha$ & 6  & $0.082$ & $0.062$ & $\phantom{-}0.013$ & $-0.005$ & $0.960$ & $0.946$ \\
              & 12  & $0.058$ & $0.028$ & $\phantom{-}0.018$ & $\phantom{-}0.004$ & $0.953$ & $0.945$ \\

$\eta_\gamma$ & 6 & $0.034$ & $0.017$ & $-0.072$ & $-0.020$ & $0.963$ & $0.933$ \\
              & 12  & $0.030$ & $0.017$ & $-0.053$ & $-0.023$ & $0.969$ & $0.938$ \\

$\eta_\lambda$ & 6  & $0.057$ & $0.028$ & $-0.000$ & $\phantom{-}0.000$ & $0.942$ & $0.937$ \\
               & 12  & $0.038$ & $0.023$ & $-0.000$ & $\phantom{-}0.000$ & $0.955$ & $0.946$ \\

$\eta_\mu$ & 6 & $0.042$ & $0.032$ & $-0.002$ & $\phantom{-}0.000$ & $0.946$ & $0.944$ \\
           & 12 & $0.021$ & $0.020$ &  $-0.000$ & $\phantom{-}0.000$ & $0.945$ & $0.945$ \\

$\eta_\sigma$ & 6 & $0.003$ & $0.012$ & $\phantom{-}0.018$ & $\phantom{-}0.022$ & $0.914$ & $0.892$ \\
              & 12 & $0.002$ & $0.004$ & $\phantom{-}0.010$ & $\phantom{-}0.020$ & $0.921$ & $0.817$ \\
\hline 
\end{tabular}
    \begin{tablenotes}
      \small
      \item Results are based on 140 and 176 estimates using a median of 6 observations per subject for $n = 300$  and $n = 600$, respectively, and 190 and 197 estimates using a median of 12 observations per subject.
    \end{tablenotes}
\end{center}
\end{threeparttable}
\end{table}

In conclusion, the simulations show that both truly linear associations and truly nonlinear associations can be estimated precisely and unbiasedly with the flexible additive joint model. Estimates are comparable between \pkg{bamlss} and \pkg{JMbayes}; however, the latter shows coverage issues, especially when truly nonlinear associations are present. 
The model is further able to distinguish between nonlinear associations of different subgroups. In this rather complex case however, estimation is only stable with enough data, both regarding the total number of subjects and the number of observations per subject, and is more stable in areas of $\eta_\mu$ where more longitudinal information is available. 
Stability issues in the estimation can be alleviated by restarting the algorithm with a different seed.

\section{Application}
\label{sec5}

We illustrate the flexible modeling approach on the widely used PBC biomedical data \citep{murtaugh_primary_1994}, included in the \proglang{R} package \pkg{JMbayes}, which is concerned with the study of survival in subjects with a rare fatal liver disease. By reanalyzing this data set with  the flexible additive joint model, assumptions and modeling alternatives can be tested. In more detail we aim to assess the adequacy of the linearity assumption of the association between marker and log-hazard and are interested in the best transformation of the marker. Our framework allows us to check several transformations and base a decision on the DIC and/or residual diagnostics without having to worry about a potentially resulting nonlinear association between the transformed marker and the log-hazard. Lastly, the analysis of subgroups regarding their association between marker and log-hazard is of interest. 

In this study 312 subjects were followed in the Mayo Clinic from 1974 to 1984 to study the influence of the drug D-penicillamine on the survival of the patients. Visits were scheduled at six months, 12 months and annually thereafter. In the dataset 140 subjects died during follow-up with a median survival time of 3.72 years (IQR: 2.08, 6.66) and 172 survived of which 29 received a transplant with a median censoring time of 7.77 (IQR: 5.73, 9.91). In total there are 1945 longitudinal observations with a median number of visits per subject of 5 (IQR: 3, 9). 

To illustrate the general framework we model the survival of PBC-patients as a function of the baseline covariates medication (drug vs. placebo), age at study entry in years and the presence of an enlarged liver at baseline. We chose these baseline covariates based on previous joint model analyses of the data \citet{rizopoulos_joint_2012, R_JMbayes}. The focus of the analysis is the association between the levels of serum bilirubin, a biomarker expected to be a strong indicator of disease progression, and the log-hazard of death. To account for individual nonlinear marker trajectories we model the levels of serum bilirubin using functional random intercepts with 5 basis functions per subject. 

To further explore the influence of the marker parameterization on the association we fit three models, differing in their association between serum bilirubin and survival. First, we model serum bilirubin using the log-transformed marker $\log(Bilirubin)$, as previously used in \citep{rizopoulos_joint_2012, R_JMbayes} and allow the association to be nonlinear. Second, we use a square-root transformation of the raw marker values $\sqrt{Bilirubin}$ and again allow the association to be nonlinear. Third, we allow the non-linear association between $\log(Bilirubin)$ and the log-hazard to also vary between the patients with an enlarged liver at baseline and those without. This predictor $\eta_{\alpha}$ is parameterized as potentially nonlinear effect for both groups, subject to the sum-to-zero constraint as explained in Section \ref{identifiability}, with an additional intercept for the group of subjects with an enlarged liver to allow not only for differences in the nonlinearity of the biomarker effect but also in the overall level. As the group difference for the hazard is already included in $\eta_\alpha$, the baseline effect of an enlarged liver is not included in $\eta_\gamma$ in model 3 to avoid redundancy. 
As our focus lies primarily on the association between the biomarker and survival, and to avoid instabilities in the estimation, we censor subjects 1 year after their last longitudinal measurement. 
In all three models, and in line with previous analyses, the treatment is not associated with survival (log-hazard effect estimate [95\% credibility interval]: model 1: -0.03 [-0.42; 0.34]; model 2: -0.02 [-0.42; 0.36]; model 3: -0.01 [-0.39; 0.39]) whereas age at baseline is positively associated with the hazard of death (model 1: 0.05 [0.03; 0.07]; model 2: 0.05 [0.04; 0.07]; model 3: 0.05 [0.03; 0.07]). Additionally subjects with an enlarged liver at baseline have a higher risk of dying in model 1 (0.76 [0.29; 1.21]) as well as in model 2 (0.77 [0.32; 1.21]). In model 3 this effect is included in the group-specific intercept for the association where the nonlinear effect of the predicted marker $\eta_\mu$ in the two subgroups is subject to the identifiability constraint explained in Section \ref{identifiability}. Under this parameterization subjects with an enlarged liver at baseline have a higher log-hazard for the event (0.49 [-0.36; 1.45]) with the respective credibility interval covering 0. Note that due to the identifiability constraint applied to the nonlinear terms in the association $\eta_\alpha$ this group-specific intercept is not directly comparable to the group effect in models 1 and 2.

\begin{figure}[htb] 
  \subfloat[\label{fig:app_m1}]{%
    \includegraphics[width=0.3\textwidth]{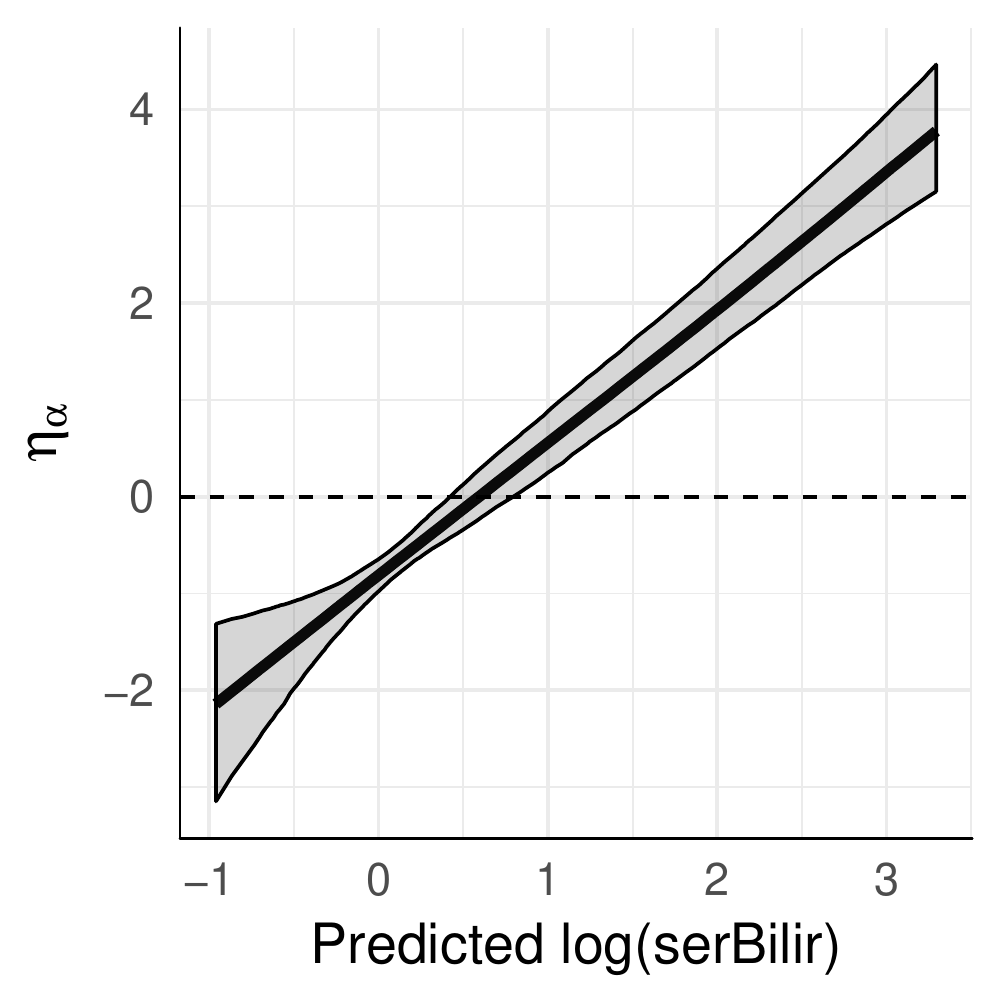} 
  } 
  \subfloat[\label{fig:app_m2}]{%
    \includegraphics[width=0.3\textwidth]{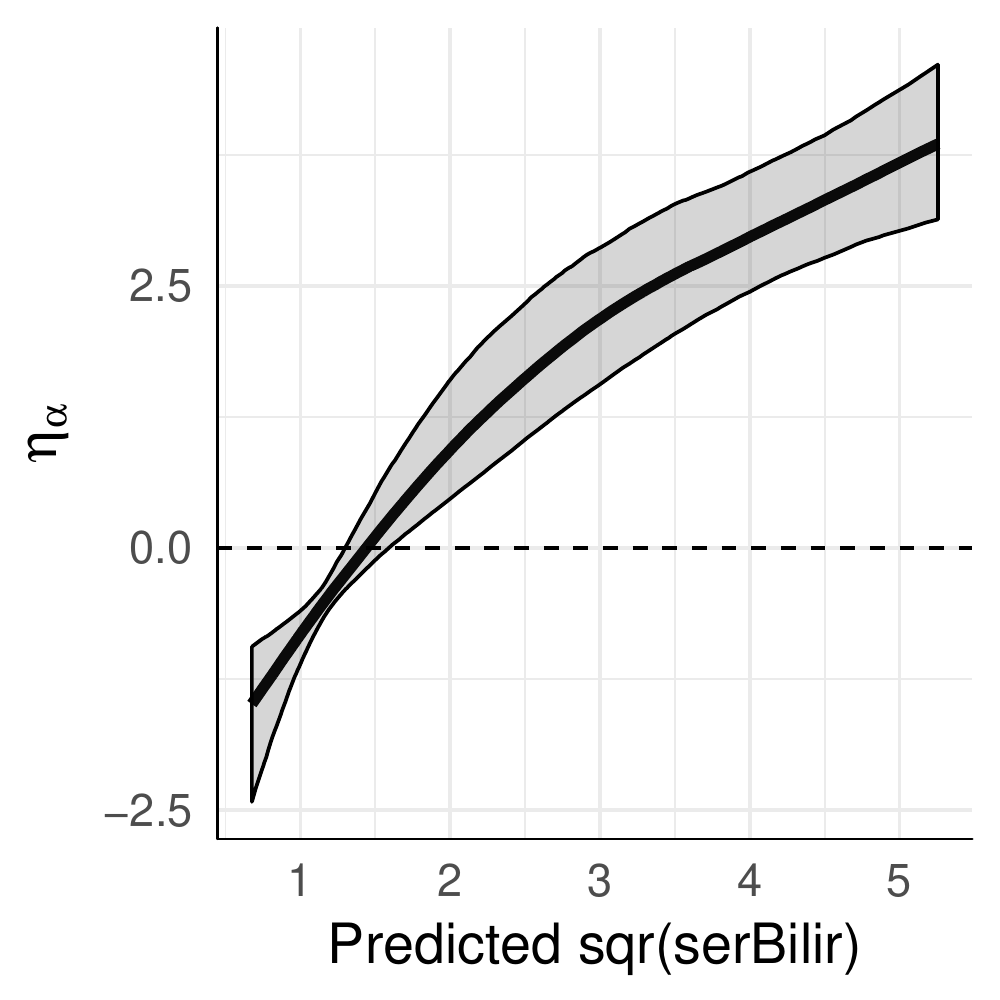} 
  } 
    \subfloat[\label{fig:app_m3}]{%
    \includegraphics[width=0.39\textwidth]{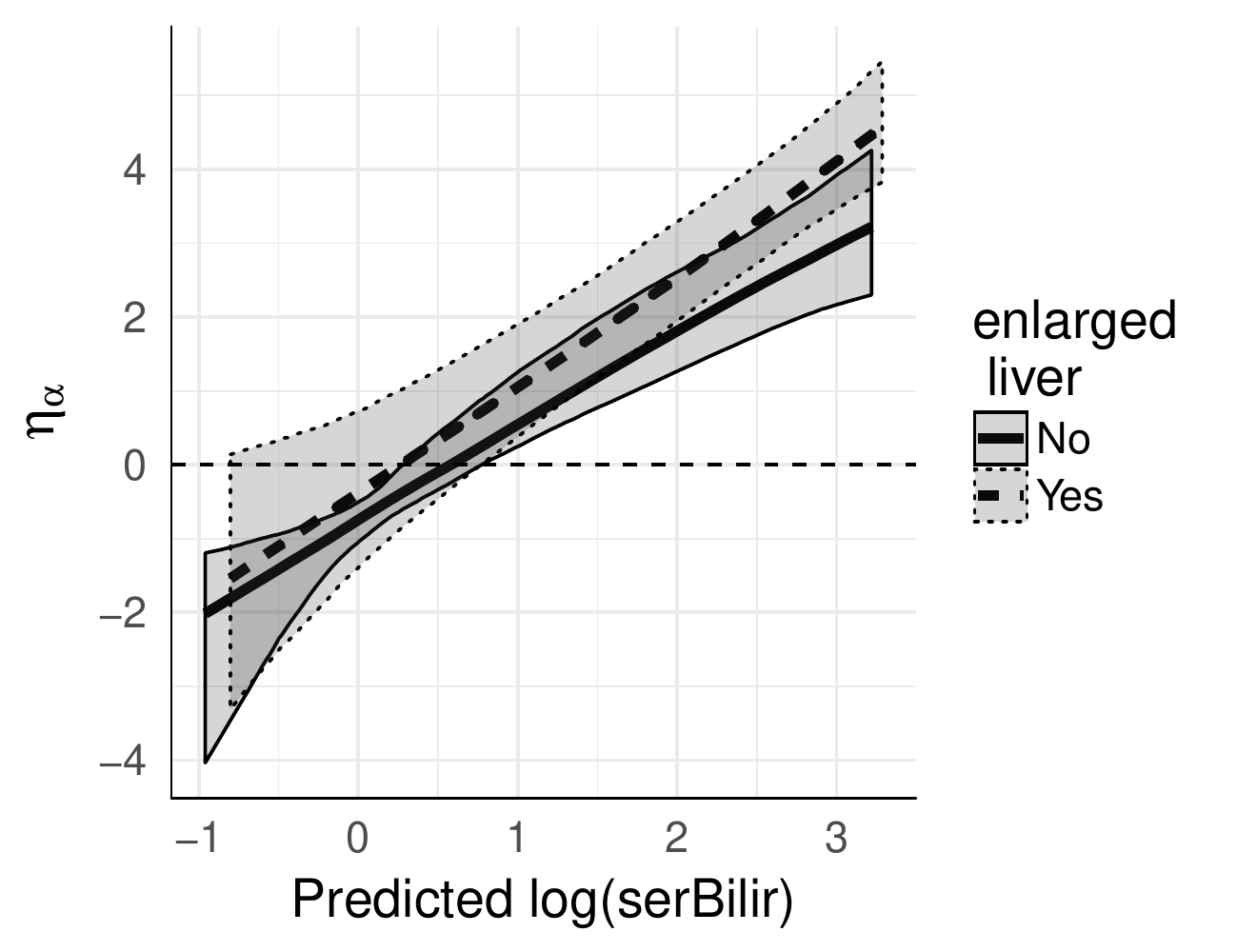} 
  } 
  \caption{Estimated posterior mean and credibility band for the association  $\eta_{\alpha}(\eta_\mu(t))$ in the PBC data. (a) model 1: nonlinear effect of  $\log(Bilirubin)$ (b) model 2: nonlinear effect of  $\sqrt{Bilirubin}$ (c) model 3: nonlinear effect of  $\log(Bilirubin)$ of patients with and without enlarged liver at baseline.}
\label{fig:app_effects}
\end{figure}

The focus of interest is the nonlinearly modeled association predictor $\eta_{\alpha}$. As Figure \ref{fig:app_effects} shows, the association between marker and the log-hazard for the event is linear when using the log-transformed marker $\log(Bilirubin)$ and nonlinear when transformed differently as $\sqrt{Bilirubin}$. In model 3 the groups differ in their overall level, although the credibility interval of the group intercept covers 0. Additionally the slope of the association is highly similar in both groups. When comparing the models via DIC, model 1 achieves the lowest DIC (1876.76) followed by model 3 (1889.67) and 2 (2194.58). 

Traceplots of the estimated coefficients $\bm{\beta}_\alpha$ as well as results from sensitivity analyses using different priors for the variance parameters (differently specified IG and Half-Cauchy hyperpriors), showing robustness of the results, can be found in the Supplementary Information.

Our flexible joint model thus allowed us to check previously made model assumptions for this data set and to conclude that in this particular case, a linear association that is not covariate-dependent is sufficient to model the relationship between the log-marker and the log-hazard. Additionally, nonlinear associations can also be captured in real data if necessary, as shown for the square-root transformation in model 2. The model potentially further allows to observe group-specific nonlinear association structures for subgroups of subjects, even though no strong group structure was present in this data set.

\section{Discussion and Outlook}
\label{sec6}
In this work a highly flexible additive joint model is presented, which allows for nonlinear, potentially covariate-dependent association structures between marker values and the log-hazard of an event. The benefits and challenges of this flexibility were shown based not only on simulated data but also on the well-known PBC data set. 

Using this new model the generally unchecked linearity assumption as well as the appropriateness of transformations of marker values can be assessed in the context of joint models. This is particularly important if marker values need to be transformed to better fulfill the normality assumption in the longitudinal submodel and different transformations are compared. It is clear that several transformations cannot fulfill the linearity and normality assumption simultaneously and relaxing the linearity assumption allows to choose the most appropriate model in terms of residual normality and/or DIC. The modeling of nonlinear associations between a longitudinal marker and the log-hazard does not only avoid bias but also allows further insights into underlying disease mechanisms. Additionally, subgroups of subjects with different marker associations can be identified. The simulation results show that our model can identify truly linear as well as truly nonlinear associations. We used the model to check the linearity assumption when using transformed Bilirubin values in the PBC data set and could confirm that the association is linear if $\log(Bilirubin)$ is used, while using $\sqrt{Bilirubin}$ would necessitate estimating a nonlinear association structure. 

This flexible modeling however also comes at a price. When modeling longitudinal trajectories using flexible functional random intercepts and allowing for nonlinear association structures, many subjects and a relatively dense grid of measurements until the event time are necessary in order to achieve a stable estimation. Further, there should not be large gaps between the latest longitudinal measurements and the event time to allow for a stable estimation. If these gaps are present in real data, censoring as in Section \ref{sec5} can alleviate the stability issue. Additionally the estimation takes more time than standard joint models but can be parallelized if corresponding computing facilities are available.

Future work could investigate further numerical approaches to stabilize estimation for complex model specifications with relatively small datasets. In addition, we plan to implement additional nonlinear association structures within the \proglang{R} package \pkg{bamlss} and to speed up the computations further in order to allow for a broader usage of this flexible additive joint model framework in practice.

\newpage

\section*{Appendix}
\label{sec8}

\subsection{Setup of model structures}

The predictor vectors $\bm{\eta}_k$, function evaluations $\bm{f}_{k}$ and design matrices $\bcx_{k}$ take on different forms for the different predictors  $k \in \{\gamma, \lambda, \alpha, \mu, \sigma \}$ in the survival and longitudinal submodel. Note that we drop the subscript $m$ in the following for ease of notation. The following Table \ref{tbl:dimensions} gives an overview.

\begin{table}[H]
\begin{center}
\caption{Overview of the predictor vectors, function evaluations and design matrices in the survival and longitudinal submodel.}
\label{tbl:dimensions}
\begin{tabular}{lllll}
\hline
                    & predictor vector                                                                  & function evaluation                                                                                                  & design matrix                                                         \\ \hline
survival model     &  & & & \\
$k \in \{\gamma\}$ & $\bm{\eta}_k = [\eta_{ki}]^{\top}$                             & $\bm{f}_{k}=[f_{k}(\bm{\tilde{x}}_{k i})]^{\top}$                                                & $\bcx_{k}$ \\
& $n \times 1$  & $n \times 1$  & $n \times p_{k}$ \\
$k \in \{\lambda, \alpha, \mu\}$ & $\bm{\eta}_k(t)= [\eta_{ki}(t)]^{\top}$                     & $\bm{f}_{k}(t)=[f_{k}(\bm{\tilde{x}}_{k i}(t), t)]^{\top}$                                 & $\bcx_k(t)$ \\
& $n \times 1$  & $n \times 1$  & $n \times p_{k}$ \\
longitudinal model & & & & \\
$k \in \{\mu, \sigma \}$         &$\bm{\eta}_k(\bt)= [\bm{\eta}_{ki}(\bt_i)^{\top}]^{\top}$ & $\bm{f}_{k}(\bt)=[f_{k}(\bm{\tilde{x}}_{k i}(\bt_i), \bt_i)^{\top}]^{\top}$ & $\bcx_{k}(\bt)$     \\
 & $N \times 1$  & $N \times 1$  & $N \times p_{k}$    \\ \hline                                                  
\end{tabular}
    \begin{tablenotes}
      \small
      \item For ease of notation we denote the vector $\ba^\top = [a_1, \cdots , a_n]$ as $[a_i]$ for $i = 1, \dots ,n$ and drop the subscript $m$ for the different terms per predictor in this illustration.
    \end{tablenotes}
\end{center}
\end{table}

For the computation of likelihood, score vector and Hessian, evaluations of the predictors are also necessary at  the survival times $\bct$. Here, $\bcx_k(\bct)$ denotes the respective $n \times p_k$ design matrix of evaluations of the time-varying predictors of the survival part $k \in \{\lambda, \alpha, \mu\}$ at time points $\bct$.

\subsection{Likelihood, Scores, and Hessian}

In the following score vectors and Hessians for the regression coefficients of every predictor are presented. Please note that in comparison with the previously presented flexible additive joint model \citep{kohler_flexible_2017} only the score and Hessians for the predictors $\eta_\alpha$ and $\eta_\mu$ have changed relevantly for the nonlinear specification.  The full log-likelihood is
\begin{align}
\ell\left[\bm{\theta} | \bct, \bm{\delta}, \bm{y}\right] = &  \bm{\delta}^{\top}\left[\bcx_{\lambda}(\bct)\bm{\beta}_{\lambda} + \bcx_{\gamma}\bm{\beta}_{\gamma} + \left[\bm{g}_1\left(\bcx_{\mu}(\bct)\bm{\beta}_{\mu}\right) \odot \bm{g}_2(\tilde{\bcx}_{\alpha}(\bct)) \right]\bm{\beta}_{\alpha} \right] \\
& -\sum_{i=1}^{n}\exp\left(\bm{x}_{\gamma i}^{\top}\bm{\beta}_{\gamma}\right)\int_{0}^{T_{i}}\exp\left[\bm{x}_{\lambda i}^{\top}\left(u\right)\bm{\beta}_{\lambda}+\left[\bm{g}_1\left(\bm{x}_{\mu i}^{\top}\left(u\right)\bm{\beta}_{\mu}\right) \odot \bm{g}_2(\bm{\tilde{x}}_{\alpha i}^{\top}(u))\right]\bm{\beta}_{\alpha}\right] \ du \\ 
& -\frac{N}{2}\log(2\pi)- \bm{1}_{N}^{\top} \bcx_{\sigma}\left(\bt\right)\bm{\beta}_{\sigma} -\frac{1}{2}(\bm{y}-\bcx_{\mu}\left(\bt\right)\bm{\beta}_{\mu})^{\top} \bm{R}^{-1}(\bm{y}-\bcx_{\mu}\left(\bt\right)\bm{\beta}_{\mu}).
\end{align}
For the flexible association in \eqref{general} the term $\bm{g}_1\left(\bcx_{\mu}(\bct)\bm{\beta}_{\mu}\right)$ reduces to $\bcx_{\mu}(\bct)\bm{\beta}_{\mu}$ for a linear association and is $\bcb\left(\bcx_{\mu}(\bct)\bm{\beta}_{\mu}\right)$ for a nonlinear association. Likewise the term $\bm{g}_2(\tilde{\bcx}_{\alpha}(\bct))$ reduces to $\bf{1}_{n}$ for a simple constant association, is the covariate vector or design matrix of the parametric input for covariate-dependent associations and is the evaluation of a spline basis matrix for a time-varying association. We denote this term in the following as $\bcx_{\alpha 2}$ to represent all three possible forms. \\
The resulting log-posterior is
\begin{equation}
\begin{split}
\log p(\bm{\theta} | \bct, \bm{\delta}, \bm{y}) \propto \ell\left[\bm{\theta} | \bct, \bm{\delta}, \bm{y}\right]
+ \sum_{k \in \{\lambda, \gamma, \alpha, \mu, \sigma\} } \sum_{m=1}^{M_k} \left[\log p(\bm{\beta}_{km}|\bm{\tau}_{km}^2) + \log p(\bm{\tau}_{km}^2)\right].
\end{split}
\end{equation}

The scores $\bm{s}(\bm{\beta}_{k})$ and Hessians $\bm{H}(\bm{\beta}_{k})$ are computed as the sum of the respective derivatives of the log-likelihood and of the log-prior densities. The latter are for example $-\frac{1}{\tau_{km}^2} \bck_{km} \bm{\beta}_{km}$ and $-\frac{1}{\tau_{km}^2} \bck_{km}$ for the multivariate normal prior as specified in Section \ref{sec3}. The score vectors $\bm{s}^\ast(\bm{\beta}_{k})$ and Hessians $\bm{H}^\ast(\bm{\beta}_{k})$ of the log-likelihood function are presented in the following.

\subsubsection{Score Vectors}
\label{Scores}

\begin{align}
\bm{s}^\ast(\bm{\beta}_{\mu}) = \frac{\partial\ell}{\partial\bm{\beta}_{\mu}} 
 = &  \bcx_{\mu}\left(\bt\right)^{\top}\bm{R}^{-1}\left(\bm{y}-\bcx_{\mu}\left(\bt\right)\bm{\beta}_{\mu}\right)+ \bcx_{\mu}^{\top}\left(\bct\right)\diag(\bm{\delta})\left[\bm{g}_1^\prime\left(\bcx_{\mu}(\bct)\bm{\beta}_{\mu}\right) \odot \bcx_{\alpha 2}(\bct)\right]\bm{\beta}_{\alpha}\\
& -\sum_{i=1}^{n}\exp\left(\bm{x}_{\gamma i}^{\top}\bm{\beta}_{\gamma}\right)\int_{0}^{T_{i}}\psi_i(u) \  \left[\bm{g}_1^{\prime}\left(\bm{x}_{\mu i}^{\top}\left(u\right)\bm{\beta}_{\mu}\right) \odot \bx_{\alpha 2 i}^{\top}(u) \right] \bm{\beta}_{\alpha}\bm{x}_{\mu i}(u)du\\
\bm{s}^\ast(\bm{\beta}_{\alpha}) = \frac{\partial\ell}{\partial\bm{\beta}_{\alpha}} 
 = & \bm{\delta}^{\top} \left[\bm{g}_1\left(\bcx_{\mu}\left(\bct\right)\bm{\beta}_{\mu}\right) \odot \bcx_{\alpha 2}(\bct) \right] \\
 & -\sum_{i=1}^{n}\exp\left(\bm{x}_{\gamma i}^{\top}\bm{\beta}_{\gamma}\right)\int_{0}^{T_{i}}\psi_i(u) \ \left[\bm{g}_1\left(\bm{x}_{\mu i}^{\top}\left(u\right)\bm{\beta}_{\mu}\right) \odot \bx_{\alpha 2 i}^{\top}(u) \right]^\top du\\
\bm{s}^\ast(\bm{\beta}_{\gamma})  = \frac{\partial\ell}{\partial\bm{\beta}_{\gamma}} 
 = & \bm{\delta}^{\top}\bcx_{\gamma}-\sum_{i=1}^{n}\exp\left(\bm{x}_{\gamma i}^{\top}\bm{\beta}_{\gamma}\right)\bm{x}_{\gamma i}\int_{0}^{T_{i}}\psi_i(u) \ du\\
  \bm{s}^\ast(\bm{\beta}_{\lambda}) = \frac{\partial\ell}{\partial\bm{\beta}_{\lambda}} 
 = & \bm{\delta}^{\top}\bcx_{\lambda}\left(\bct\right)-\sum_{i=1}^{n}\exp\left(\bm{x}_{\gamma i}^{\top}\bm{\beta}_{\gamma}\right)\int_{0}^{T_{i}}\psi_i(u) \ \bm{x}_{\lambda i}\left(u\right)du\\
  \bm{s}^\ast(\bm{\beta}_{\sigma}) = \frac{\partial\ell}{\partial\bm{\beta}_{\sigma}} 
 = & - \bcx_{\sigma}\left(\bt\right)^{\top} \bm{1}_{N} + \left[\bcx_{\sigma}\left(\bt\right) \odot \left(\bm{y}-\bcx_{\mu}\left(\bt\right)\bm{\beta}_{\mu}\right)\right]^{\top}
\bm{R}^{-1}\left(\bm{y}-\bcx_{\mu}\left(\bt\right)\bm{\beta}_{\mu}\right)\\
\end{align}
with  $\psi_i(u) = \exp\left[\bm{x}_{\lambda i}^{\top}\left(u\right)\bm{\beta}_{\lambda}+\left[\bm{g}_1\left(\bm{x}_{\mu i}^{\top}\left(u\right)\bm{\beta}_{\mu}\right)^{\top} \odot \bx_{\alpha 2 i}^{\top}(u) \right]\bm{\beta}_{\alpha}\right]$ and  the diagonal matrix $\bm{R}=\diag\left(\exp\left[\bcx_{\sigma}\left(\bt\right)\bm{\beta}_{\sigma}\right]^2\right)$. 
For the score vector $\bm{s}^\ast(\bm{\beta}_{\mu})$ the derivative of $\bm{g}_1\left(\bm{x}^\top_{\mu i}(u) \bm{\beta}_{\mu}\right)$ with respect to $\bm{\beta}_{\mu}$ is needed which can be derived by chain rule 
\begin{equation}
\frac{\partial \bm{g}_1(\bm{x}^\top_{\mu i}(u) \bm{\beta}_{\mu})}{\partial \bm{\beta}_{\mu}} =
 \frac{\partial \bm{g}_1(\bm{\eta}_{\mu}(u))}{\partial \bm{\eta}_{\mu}(u)} \cdot \frac{\partial \bm{\eta}_{\mu}(u)}{\partial \bm{\beta}_{\mu}}.
\end{equation}
The derivative of $\bm{g}_1\left(\bcx_{\mu}(\bct) \bm{\beta}_{\mu}\right)$ follows analogously. Whereas the inner derivative $\frac{\partial \bm{\eta}_{\mu}(u)}{\partial \bm{\beta}_{\mu}} = \bm{x}(u)$ is the same for both linear and nonlinear associations, the outer derivative, which we denote by $\bm{g}_1^{\prime}\left(\bm{x}_{\mu i}^\top(u) \bm{\beta}_{\mu}\right)$, differs between the parameterizations. For linear associations it holds that $\bm{g}_1^{\prime}\left(\bm{x}_{\mu i}^\top(u) \bm{\beta}_{\mu}\right) = 1$ and $\bm{g}_1^{\prime}\left(\bcx_{\mu}(\bct) \bm{\beta}_{\mu}\right) = \bf{1}^\top_n$. Nonlinear associations as implemented using penalized B-splines in \pkg{bamlss} yield $\bm{g}_1^{\prime}\left(\bm{x}^\top_{\mu i}(u) \bm{\beta}_{\mu}\right) =\bcb^{\prime}(\bm{x}^\top_{\mu i}(u) \bm{\beta}_{\mu})$ 
and $\bm{g}_1^{\prime}\left(\bcx_{\mu}(\bct) \bm{\beta}_{\mu}\right) =\bcb^{\prime}(\bcx_{\mu}(\bct) \bm{\beta}_{\mu})$, which have a straightforward analytical solution for the derivative \citep{fahrmeir_regression}
\begin{equation}
\frac{\partial}{\partial z} \sum_d B_d^l(z) = l \left(\frac{1}{\kappa_{d} - \kappa_{d-1}} B_{d-1}^{l-1}(z) - \frac{1}{\kappa_{d+1} - \kappa_{d+1-l}}B_{d}^{l-1}(z)\right),
\end{equation}
where $l$ denotes the degree of the spline, $d$ is the index for the basis functions  and $\kappa$ denotes the knots with the interior knots $\kappa_1, \dots, \kappa_m$ and $2l$ outer knots.

\subsubsection{Hessian}
\begin{align}
\bm{H}^\ast(\bm{\beta}_{\mu}) =\frac{\partial^{2}\ell}{\partial\bm{\beta}_{\mu}\partial\bm{\beta}_{\mu\top}}
 = & 
-\bcx_{\mu}\left(\bt\right)^{\top}\bm{R}^{-1}\bcx_{\mu}\left(\bt\right) + \bcx_{\mu}^{\top}\left(\bct\right)\diag(\bm{\delta})\left[\bm{g}_1^{\prime \prime}\left(\bcx_{\mu}(\bct)\bm{\beta}_{\mu}\right) \odot \bcx_{\alpha 2}(\bct)\right]\bm{\beta}_{\alpha} \bcx_{\mu}\left(\bct\right)\\
& -\sum_{i=1}^{n}\exp\left(\bm{x}_{\gamma i}^{\top}\bm{\beta}_{\gamma}\right)\int_{0}^{T_{i}}\psi_i(u) \cdot \\
&  \left[\left(\left[\bm{g}_1^{\prime}\left(\bm{x}_{\mu i}^{\top}\left(u\right)\bm{\beta}_{\mu}\right) \odot \bx_{\alpha 2 i}^{\top}(u)\right]  \bm{\beta}_{\alpha}\right)^2 + \ \left[\bm{g}_1^{\prime \prime}\left(\bm{x}_{\mu i}^{\top}\left(u\right)\bm{\beta}_{\mu}\right) \odot \bx_{\alpha 2 i}^{\top}(u)\right]  \bm{\beta}_{\alpha} \right] \cdot \\
& \bm{x}_{\mu i}(u)\bm{x}^{\top}_{\mu i}(u) du\\
\bm{H}^\ast(\bm{\beta}_{\alpha}) = \frac{\partial^{2}\ell}{\partial\bm{\beta}_{\alpha}\partial\bm{\beta}_{\alpha}^{\top}} 
 = & -\sum_{i=1}^{n}\exp\left(\bm{x}_{\gamma i}^{\top}\bm{\beta}_{\gamma}\right)\int_{0}^{T_{i}}\psi_i(u) \ \left[\bm{g}_1\left(\bm{x}_{\mu i}^{\top}\left(u\right)\bm{\beta}_{\mu}\right) \odot \bx_{\alpha 2 i}^{\top}(u)\right] \cdot \\
 & \left[\bm{g}_1\left(\bm{x}_{\mu i}^{\top}\left(u\right)\bm{\beta}_{\mu}\right) \odot \bx_{\alpha 2 i}^{\top}(u)\right]^\top du\\ 
\bm{H}^\ast(\bm{\beta}_{\gamma})  = \frac{\partial^{2}\ell}{\partial\bm{\beta}_{\gamma}\partial\bm{\beta}_{\gamma}^{\top}} 
 = & 
-\sum_{i=1}^{n}\exp\left(\bm{x}_{\gamma i}^{\top}\bm{\beta}_{\gamma}\right)\bm{x}_{\gamma i}\bm{x}_{\gamma i}^{\top}\int_{0}^{T_{i}}\psi_i(u) \ du\\
\bm{H}^\ast(\bm{\beta}_{\lambda}) = \frac{\partial^{2}\ell_{i}}{\partial\bm{\beta}_{\lambda}\partial\bm{\beta}_{\lambda}^{\top}} 
 = & -\sum_{i=1}^{n}\exp\left(\bm{x}_{\gamma i}^{\top}\bm{\beta}_{\gamma}\right)\int_{0}^{T_{i}}\psi_i(u) \ \bm{x}_{\lambda i}\left(u\right)\bm{x}_{\lambda i}^{\top}\left(u\right)du\\
\bm{H}^\ast(\bm{\beta}_{\sigma}) = \frac{\partial^{2}\ell}{\partial\bm{\beta}_{\sigma}\partial\bm{\beta}_{\sigma}^{\top} } 
 = & -2 \left[\bcx_{\sigma}\left(\bt\right) \odot \left(\bm{y}-\bcx_{\mu}\left(\bt\right)\bm{\beta}_{\mu}\right)\right]^{\top}
\bm{R}^{-1}\left[\bcx_{\sigma}\left(\bt\right) \odot \left(\bm{y}-\bcx_{\mu}\left(\bt\right)\bm{\beta}_{\mu}\right)\right]\\
\end{align}
Here $\bm{g}_1^{\prime \prime}\left(\bcx_{\mu}(\bct) \bm{\beta}_{\mu}\right)$  denote the second derivatives with respect to $\bm{\eta}_{\mu}(\bct)$, i.e. the second outer derivative, which is $\bf{0}_n$ for a linear association and $B^{\prime \prime}(\bcx_{\mu}(\bct) \bm{\beta}_{\mu})$ for a nonlinear association, for which again an analytical formula exists. The same setup holds for $\bm{g}_1^{\prime \prime}\left(\bm{x}^\top_{\mu i}(u) \bm{\beta}_{\mu}\right)$.

\section*{Acknowledgements}
This work is part of Meike K\"ohler's PhD thesis within the graduate school HELENA at the Helmholtz
Zentrum M\"unchen in collaboration with the Ludwig-Maximilians-Universit\"at M\"unchen, Germany, which was supported by funds from the Helmholtz International Research Group [HIRG-0018].

\newpage
\bibliographystyle{biometrical}
\bibliography{paper_2}

\end{document}